\documentclass[prd,aps,twocolumn,a4paper,floatfix,nofootinbib,preprintnumbers]{revtex4-2}

\usepackage[utf8]{inputenc}
\usepackage{graphicx,psfrag}
\usepackage{mathrsfs}
\usepackage{amsmath,amsfonts,amssymb}
\usepackage{multirow}
\usepackage{diagbox}
\usepackage{comment}
\usepackage{enumerate}
\usepackage{booktabs}
\usepackage{bbm}
\usepackage[normalem]{ulem}
\usepackage[dvipsnames]{xcolor}
\usepackage{ulem}

\usepackage{hyperref}
\hypersetup{
    colorlinks = true,
    linkcolor = {blue},
    citecolor = {blue},
    urlcolor = {blue},
    linkbordercolor = {white},
    }

\usepackage{color}
\definecolor{cyan}{rgb}{0,0.9,0.9}
\definecolor{orange}{rgb}{0.9,0.5,0}
\definecolor{magenta}{rgb}{1,0,1}
\definecolor{purple}{rgb}{0.8,0.4,0.8}
\definecolor{gray}{rgb}{0.8242,0.8242,0.8242}
\definecolor{green}{rgb}{0.,0.8,0.}

\begin{document}
\preprint{LA-UR-22-31740}

\title{Reverse phase transitions in binary neutron-star systems with 
exotic-matter cores}

\author{Maximiliano \surname{Ujevic}$^{1}$}
\author{Henrique \surname{Gieg}$^{1}$}
\author{Federico \surname{Schianchi}$^{3}$}
\author{Swami Vivekanandji \surname{Chaurasia}$^{4}$}
\author{Ingo \surname{Tews}$^{2}$}
\author{Tim \surname{Dietrich}$^{3,5}$}

\affiliation{${}^1$ Centro de Ci\^encias Naturais e Humanas, Universidade Federal do ABC, 09210-170, Santo Andr\'e, S\~ao Paulo, Brazil}
\affiliation{${}^2$ Theoretical Division, Los Alamos National Laboratory, Los Alamos, NM 87545, USA}
\affiliation{${}^3$ Institute for Physics and Astronomy, University of Potsdam, Haus 28, Karl-Liebknecht-Str. 24/25, 14476, Potsdam, Germany}
\affiliation{${}^4$ The Oskar Klein Centre, Department of Astronomy, Stockholm University, AlbaNova, SE-10691 Stockholm,
Sweden}
\affiliation{${}^5$ Max Planck Institute for Gravitational Physics (Albert Einstein Institute), Am M\"uhlenberg 1, Potsdam 14476, Germany
}

\date{\today}

\begin{abstract}
Multi-messenger observations of binary neutron star mergers provide a unique opportunity to constrain the dense-matter equation of state. 
Although it is known from quantum chromodynamics that hadronic matter will undergo a phase transition to exotic forms of matter, e.g., quark matter, the onset density of such a phase transition cannot be computed from first principles. 
Hence, it remains an open question if such phase transitions occur inside isolated neutron stars or during binary neutron star mergers, or if they appear at even higher densities that are not realized in the Cosmos. In this article, we perform numerical-relativity simulations of neutron-star mergers and investigate scenarios in which the onset density of such a phase transition is exceeded in at least one inspiralling binary component.
Our simulations reveal that shortly before the merger it is possible that such stars undergo a ``\textit{reverse phase transition}'', i.e., densities decrease and the quark core inside the star disappears leaving a purely hadronic star at merger. 
After the merger, when densities increase once more, the phase transition occurs again and leads, in the cases considered in this work, to a rapid formation of a black hole. 
We compute the gravitational-wave signal and the mass ejection for our simulations of such scenarios and find clear signatures that are related to the post-merger phase transition, e.g., smaller ejecta masses due to the softening of the equation of state through the quark core formation.
Unfortunately, we do not find measurable imprints of the reverse phase transition. 
\end{abstract}

\maketitle

\section{Introduction}
\label{sec:intro}

Constraining the properties of cold dense nuclear matter and its equation of state (EOS) remains an open problem in modern physics.
Systematic calculations of matter properties at densities that exceed 1-2 times the nuclear saturation density, $n_{\rm sat}$, based on quantum chromodynamics (QCD) are not feasible at the moment. 
Similarly, heavy-ion collision (HIC) experiments constrain the EOS but are currently limited to the density range below $\sim 2n_{\rm sat}$~\cite{Huth:2021bsp}.
To probe dense matter beyond the densities that can be reached in HIC experiments or in nuclear theory calculation, one has to use astrophysical observations of neutron stars (NSs) and their mergers. 
The multi-messenger detection of GW170817~\cite{LIGOScientific:2017vwq,LIGOScientific:2017zic} via gravitational waves (GWs) and electromagnetic (EM) signatures has been a breakthrough that produced a wealth of information~\cite{Dietrich:2020efo}. 
Overall, the extreme densities, temperatures, fluid velocities, and stronger gravitational fields achieved during the collision of two NSs are unmatched in the Universe and provide a perfect laboratory to probe physical principles under extreme conditions. 

At merger, when gravitational fields are strongest, only numerical-relativity (NR) simulations can reliably describe the spacetime and matter fields. 
This makes NR simulations an essential tool to characterize the imprint of the binary parameters or microphysical processes, see, e.g., Refs.~\cite{Baiotti:2016qnr,Baiotti:2019sew,Dietrich:2020eud,Radice:2020ddv} for recent reviews.  
Over the years, the NR community made tremendous progress by performing high-precision simulations~\cite{Radice:2013hxh,Hotokezaka:2015xka,Kiuchi:2017pte,Bernuzzi:2016pie}, including microphysical aspects such as neutrino radiation or magnetic fields, e.g., Refs.~\cite{Anderson:2008zp,Giacomazzo:2010bx,Rezzolla:2011da,
Kiuchi:2014hja,Palenzuela:2015dqa,Dionysopoulou:2015tda,Ruiz:2017due,
Kiuchi:2017zzg,Sekiguchi:2011zd,Foucart:2017mbt,Foucart:2018gis,Foucart:2020qjb}, 
as well as investigating the impact of phase transitions on the merger dynamics~\cite{Most:2018eaw,Bauswein:2018bma,Weih:2019xvw,Blacker:2020nlq,Bauswein:2020ggy,Huang:2022mqp}. 

In fact, QCD predicts that cold matter undergoes a transition from hadronic matter at low densities to quark matter at very high energies.
However, the onset density of such a phase transition is unknown. 
Therefore, it is an open question whether isolated NSs explore such phase transitions to exotic forms of matter in their cores, e.g., Refs.~\cite{Glendenning:1992vb,Alford:2001dt,Alford:2004pf,Benic:2014jia}, which could affect the binary neutron star (BNS) system dynamics already during the inspiral~\cite{Montana:2018bkb,Gieg:2019yzq,Pang:2020ilf},
if the phase transition sets in during the post-merger phase when matter exists at higher densities~\cite{Most:2018eaw,Bauswein:2018bma,Weih:2019xvw,Blacker:2020nlq,Bauswein:2020ggy,Huang:2022mqp}, or if the phase transition appears only beyond the density regime probed in NSs.

Revealing the presence of a possible phase transition is one of the science goals of next-generation GW detectors.
This could either happen due to precise measurements of the tidal deformability of the stars during the inspiral, which is altered due to the presence of a strong first-order quark-hadron phase transition~\cite{Li:2018ayl,Christian:2018jyd,Kashyap:2021wzs,Chatziioannou:2019yko} if the onset density is sufficently low, or through post-merger GW measurements~\cite{Haque:2022dsc,Tootle:2022pvd,Fujimoto:2022xhv,Prakash:2021wpz,Liebling:2020dhf}.
Regarding the latter, numerous previous studies investigated possible shifts in the peak frequency of the post-merger GW signal when a quark-hadron phase transition is present, e.g., Refs.~\cite{Most:2018eaw,Bauswein:2018bma,Weih:2019xvw,Blacker:2020nlq,Bauswein:2020ggy,Huang:2022mqp}.
Finally, also the EM signatures might be altered due to different post-merger dynamics~\cite{Prakash:2021wpz}. 

Recently, Liebling et al.~\cite{Liebling:2020dhf} studied the effect of phase transitions in BNS systems for hadronic NSs, where a phase transition to a higher density core is found at merger time triggered by accreting matter, and a ``reverse phase transition'' is encountered in the post-merger phase due to oscillations of the central density. In contrast, in this article, we perform new NR simulations evolving BNS systems of hybrid NSs composed of a quark-matter core and a hadronic mantle. Furthermore, our simulations reveal that a ``reverse phase transition'' can happen during the inspiral shortly before the NS merger. This transition is caused by a decrease in the central mass density due to the strong deformation of the stars, which causes a change from a quark-matter core to only hadronic material before the stars come into contact.

The article is structured as follows:
We shortly review the employed methods in Sec.~\ref{sec:methods} before we present our main results in Sec.~\ref{sec:results} and conclude in Sec.~\ref{sec:conclusion}. 
Unless otherwise stated, we use units in which $G=c=M_\odot=1$.

\section{Methods and Setups}
\label{sec:methods}

\begin{table}[t]
\caption{Properties of the individual stars and BNS simulations. The columns contain the configuration name, the employed mass ratio $q=M^A/M^B\geq 1$, the gravitational masses of the individual stars $M^{A,B}$, the residual eccentricity $e$ of the BNS, the initial GW frequency $M\omega^0 _{2,2}$ of the (2,2)-mode, the Arnowitt-Deser-Misner (ADM) mass $M_\text{ADM}$, and the corresponding angular momentum $J_\text{ADM}$. 
All configurations were evolved with the resolutions of Tab.~\ref{tab:BAM_grid}.} 
\label{tab:config} 
\begin{small}
\begin{tabular}{cccccccc}
\toprule Name & $q$ & $M^A$ & $M^B$ & $e$ [$10^{-3}$] & $M\omega^0_{2,2}$ [$10^{-2}$] & $M_\text{ADM}$ & $J_\text{ADM}$ \\ \hline \hline
noPT1 & 1.00 & 1.394 & 1.394 & 1.0026 & 3.2816 & 2.7671 & 8.5150 \\ 
PT1 & 1.00 & 1.394 & 1.394 & 3.4902 & 3.2830 & 2.7671 & 8.5093 \\
noPT2 & 1.17 & 1.501 & 1.287 & 1.2720 & 3.2814 & 2.7672 & 8.4643 \\
PT2 & 1.17 & 1.501 & 1.287  & 3.7002 & 3.2805 & 2.7672 & 8.4585 \\ \hline \hline
\end{tabular}
\end{small}
\end{table}

\textbf{Simulated Configurations:} We investigate a total of four different binary configurations labeled noPT1, noPT2, PT1, and PT2. 
The configurations noPT1 and noPT2 employ EOSs without phase transition while PT1 and PT2 have phase transitions. 
The two sets 1 and 2 differ in the employed  mass ratios, 
i.e., PT1 and noPT1 represent equal mass systems with $q=M^A/M^B=1.0$, while PT2 and noPT2 are unequal mass systems with a mass ratio of $q=1.17$. 
The EOSs and system parameters are picked such that they all have the same dimensionless binary tidal deformabilities $\tilde{\Lambda}=250$. 
All simulations begin 14 to 15 orbits before the merger at an initial distance between the stars' centers of about 75~km, which leads to a dimensionless initial GW frequency of $M \omega_{2,2}^0\approx 0.032$. 
Employing the eccentricity-reduction procedure of Refs.~\cite{Moldenhauer:2014yaa,Dietrich:2015pxa}, we achieve initial eccentricities of $\mathcal{O}(10^{-3})$. 
Details about the initial configurations are listed in Tab.~\ref{tab:config}.

\begin{figure}[htpb]
\centering
  \includegraphics[width=0.85\columnwidth]{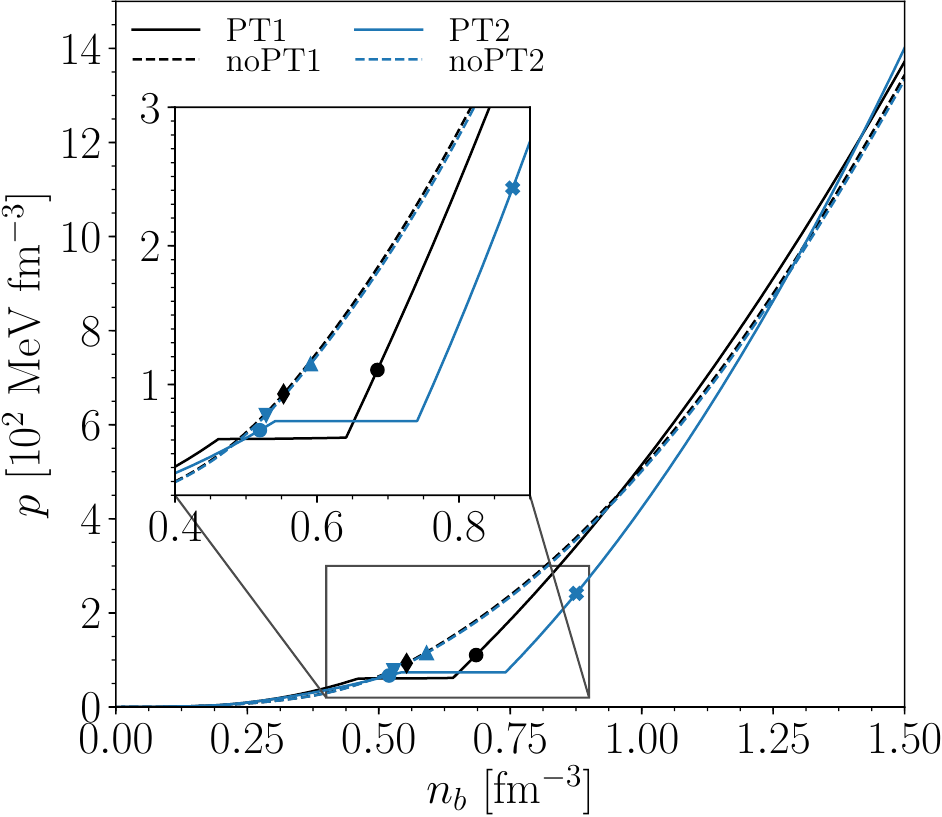}\\
  \includegraphics[width=0.85\columnwidth]{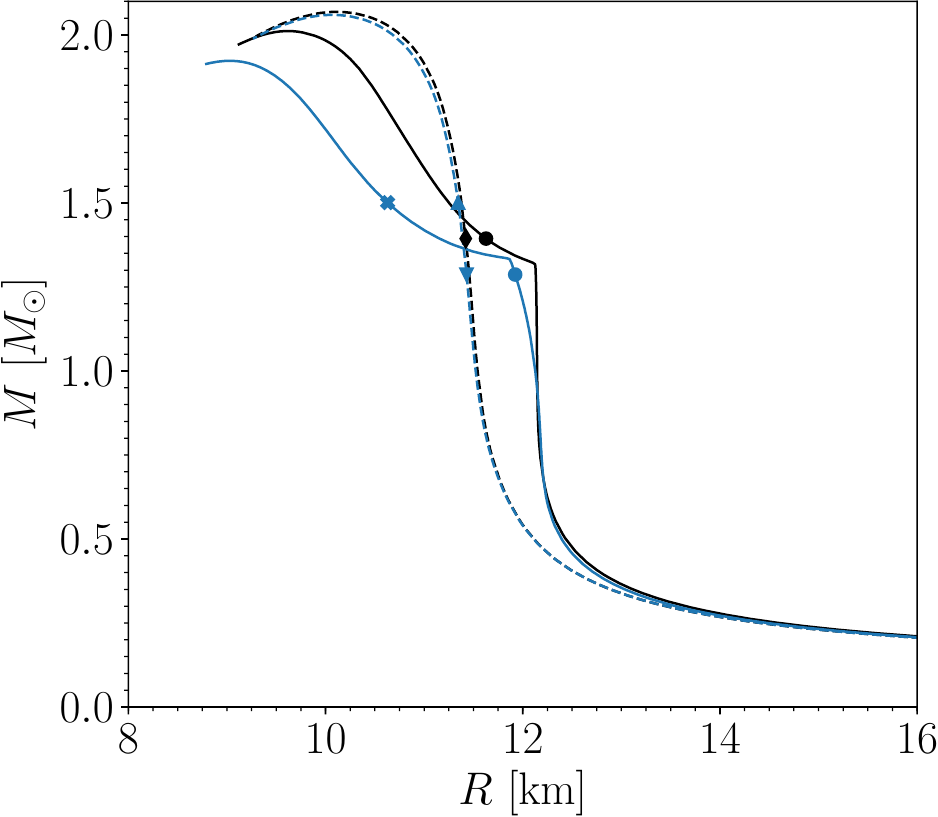}
\caption{EOSs represented by the pressure vs.\ baryon number density curves (upper panel) and mass-radius curves (bottom panel). 
The marked points in the top panel identify the central rest-mass density of the stars in the models of Tab.~\ref{tab:config}: 
PT1 (black circle), noPT1 (black diamond), PT2 star A (blue x), PT2 star B (blue circle), noPT2 star A (blue up triangle), noPT2 star B (blue down triangle). 
Masses and radii for the corresponding stars in isolation are marked accordingly in the bottom panel.} 
\label{fig:EOS_overview}
\end{figure}

\textbf{Equations of State construction:}
The EOS describes the relation between energy density $e$, pressure $p$, and temperature $T$ of dense matter and additionally depends on the composition of the system.
Given that the thermal energies are much smaller than typical Fermi energies of the particles, we can neglect temperature effects during the construction of the baseline EOSs that we use for the computation of the initial configuration, and hence, the EOS simply relates $e$ and $p$.
The EOSs used here are constrained at low densities by quantum Monte Carlo calculations using interactions derived from chiral Effective Field Theory (EFT)~\cite{Lynn:2015jua,Tews:2018kmu}. 
In particular, we employ the $V_{E\mathbbm{1}}$ interaction to constrain the EOS up to nuclear saturation density~\cite{Lynn:2015jua}.
Beyond $n_{\rm sat}$, we extrapolate the EOS using a piecewise-linear speed-of-sound extension to larger densities. 
Additionally, for two EOS we impose a first-order phase transition by setting the speed of sound within one piece to zero, with the exotic matter beyond the phase transition still being described by the speed-of-sound parametrization.
We have tuned the parameters of this extension such that the binary tidal deformabilities for all EOSs are $\tilde{\Lambda}=250$ for the mass ratios chosen here. 
For the EOSs with phase transition, the onset density and density jump have been tuned to additionally ensure that at least one of the binary NSs explores the phase transition during the inspiral. The phase transition for PT1 (PT2) sets in at baryon number density $n_b = 0.64~{\rm fm^{-3}}$  ($n_b =0.74~{\rm fm^{-3}}$), corresponding to a NS in isolation with gravitational mass $M = 1.32~M_\odot$ ($M = 1.33~M_\odot$).

Because BAM and SGRID (see the description below) did not support full tabulated EOSs at the start of this project, we have employed a piecewise-polytropic representation of the previously constructed EOSs using between 9 and 15 pieces; cf.~Appendix~\ref{app:pwp} for more details. 
Our piecewise polytropes are parametrized by the rest-mass density $\rho = m_b n_b$, where $m_b = 1.66\times10^{-24}~{\rm g}$. 
Following the methods of Ref.~\cite{Read:2008iy}, we constructed the zero-temperature part of the EOSs used in our simulations such that the pressure $p = p(\rho)$, the specific internal energy per baryon $\epsilon = \epsilon(\rho)$ and the energy density $e(\rho) = \rho(1+\epsilon(\rho))$ are continuous everywhere.

The final EOSs are represented in Fig.~\ref{fig:EOS_overview} (upper panel) and we mark the initial central density of the individual stars simulated in this work. 
The lower panel of Fig.~\ref{fig:EOS_overview} shows the corresponding mass-radius curves in which we mark the radius and masses of the individual stars. During the dynamical simulations, we extend the zero-temperature EOS by adding a thermal contribution following Ref.~\cite{Bauswein:2010dn}. 

\begin{table}[t]
\caption{Grid configurations. The columns refer to the resolution 
name, the number of levels $L$, the number of moving box levels 
$L_{\rm mv}$, the number of points in the non-moving boxes $n$, the 
number of points in the moving boxes $n_{\rm mv}$, the grid spacing in 
the finest level $h_6$ covering the neutron star, and the grid spacing in 
the coarsest level $h_0$.} \label{tab:BAM_grid}
\begin{tabular}{ccccccc}
\toprule
Name & $L$ & $L_{\rm mv}$ & $n$ & $n_{\rm mv}$ & $h_6$ & $h_0$ \\ \hline
R1 & 7 & 4 & 192 & 96 & 0.152 & 9.728 \\
R2 & 7 & 4 & 256 & 128 & 0.114 & 7.296 \\
R3 & 7 & 4 & 320 & 160 & 0.0912 & 5.8368 \\ 
R4 & 7 & 4 & 384 & 192 & 0.076 & 4.864 \\ \hline 
\hline
\end{tabular}
\end{table}

\begin{figure*}[t]
\centering
        \hspace{0.3cm}\includegraphics[width=1.92\columnwidth]{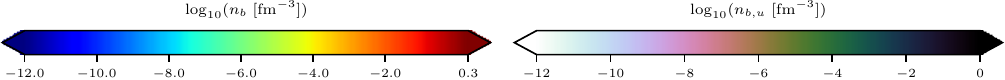}\\
        \vspace{0.3cm}
  \includegraphics[width=0.48\columnwidth]{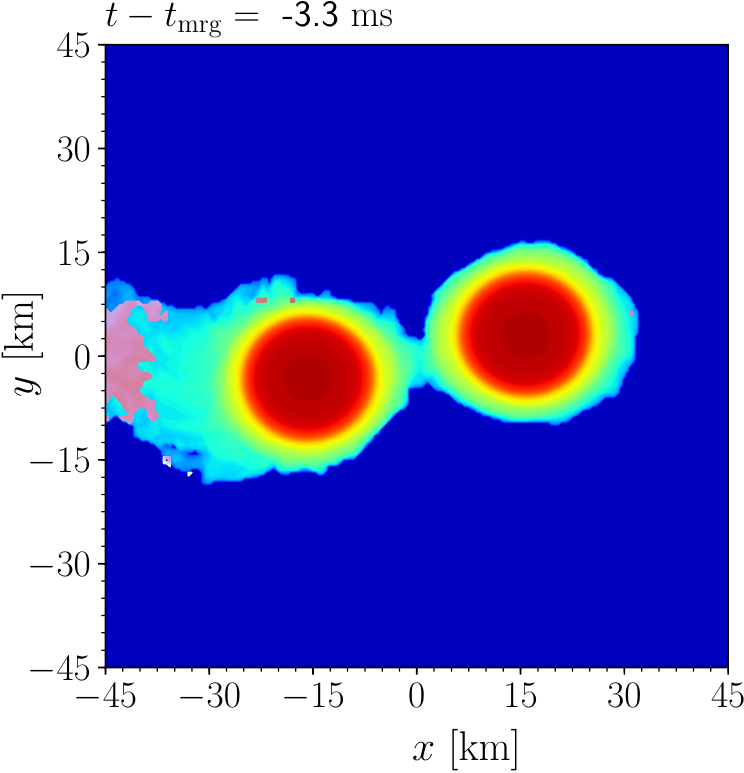} 
    \includegraphics[width=0.48\columnwidth]{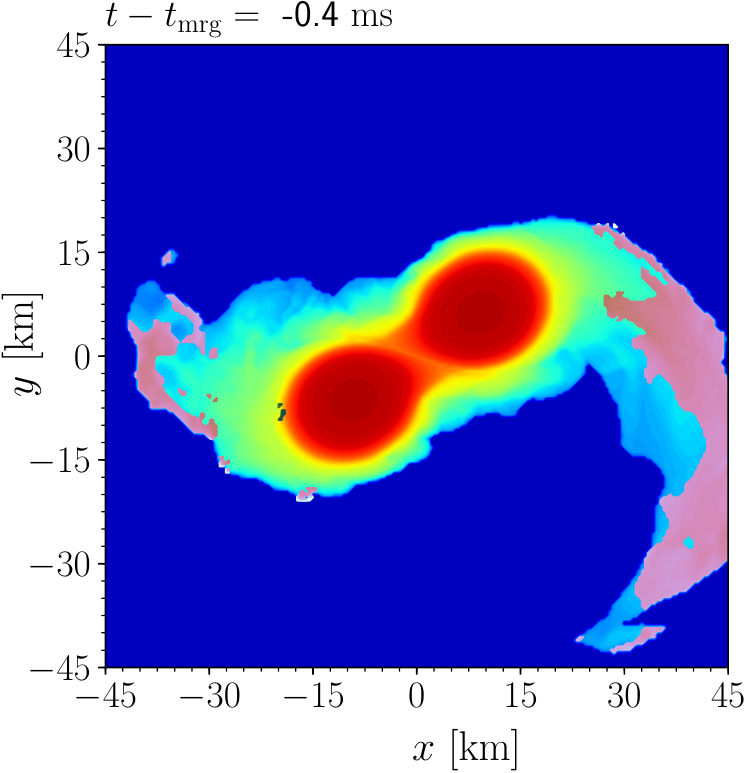} 
      \includegraphics[width=0.48\columnwidth]{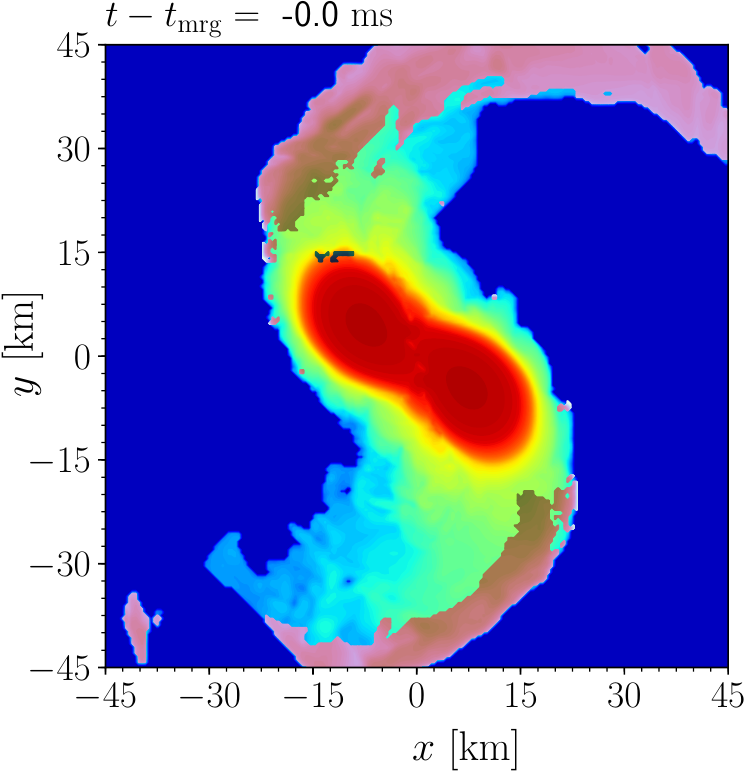} 
        \includegraphics[width=0.48\columnwidth]{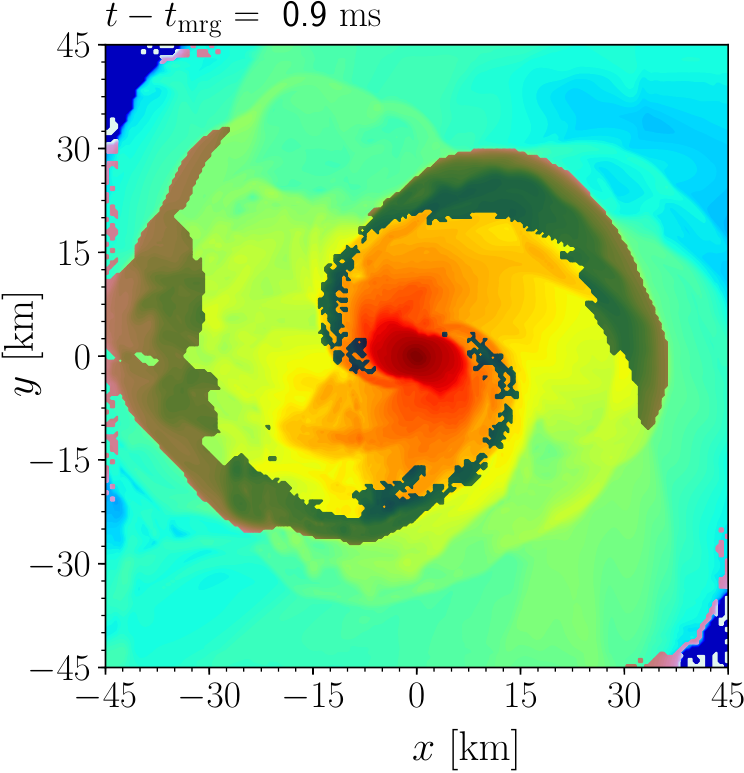} \\
	\hspace{0.05cm}\includegraphics[width=0.48\columnwidth, height=0.4\columnwidth]{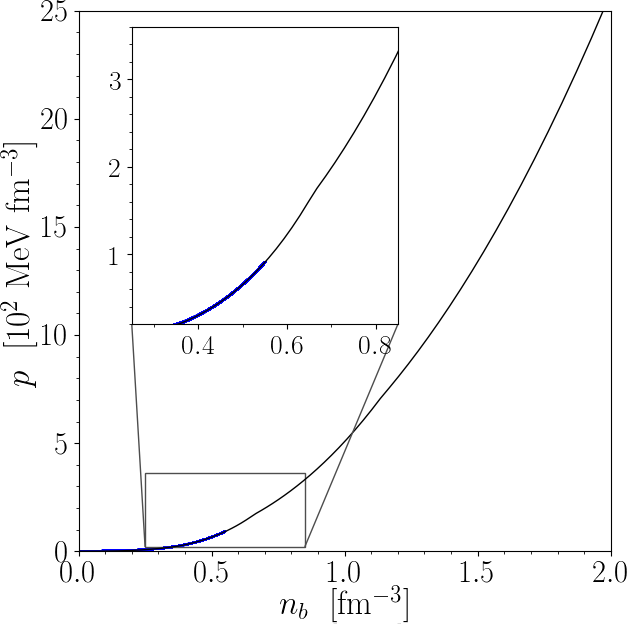}
 \includegraphics[width=0.48\columnwidth, height=0.4\columnwidth]{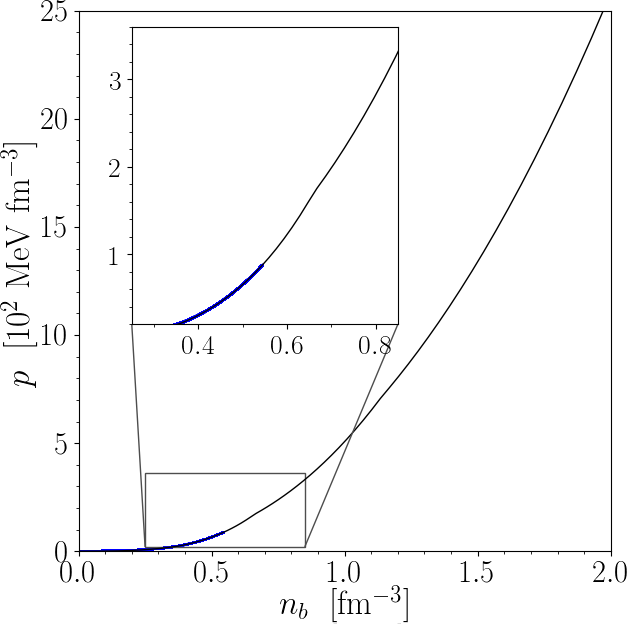}
      \includegraphics[width=0.48\columnwidth, height=0.4\columnwidth]{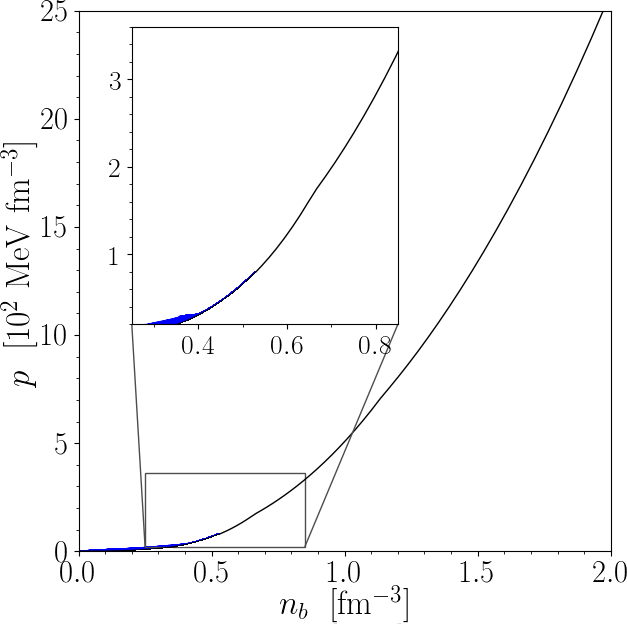}
        \includegraphics[width=0.48\columnwidth, height=0.4\columnwidth]{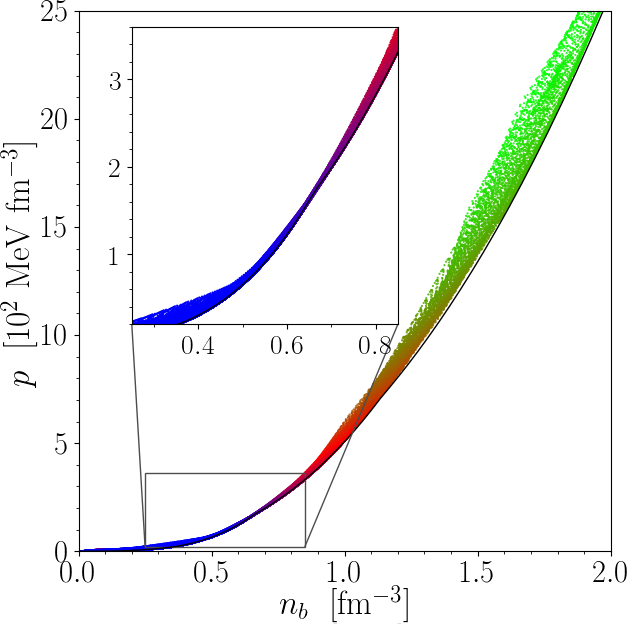}\\
                \hspace{0.3cm}\includegraphics[width=1.92\columnwidth]{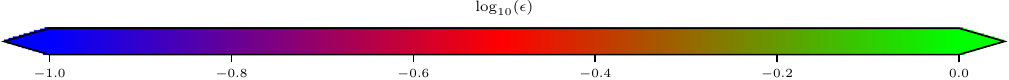}
\caption{Snapshots of the baryon number density on the equatorial plane (upper panels) and  the respective baryon number density-pressure phase diagram (lower panels), together with the zero-temperature EOS (black line) and color-coded specific internal energy $\epsilon$, are depicted for the noPT1 setup employing the highest resolution. Also, we represent the hadronic matter density $n_b$ in the upper left colorbar and the density $n_{b,u}$ of matter identified as ejecta in the upper right colorbar.} 
\label{fig:2D_noPT1}
\end{figure*}

\textbf{Numerical Methods:}
For the construction of the initial configuration, we employ the pseudospectral code SGRID~\cite{Tichy:2009yr, Tichy:2012rp, Dietrich:2015pxa, Tichy:2019ouu}, which uses the conformal thin-sandwich approach~\cite{Wilson:1995uh,Wilson:1996ty,York:1998hy} to solve the Constraint Equations. 

For the dynamical evolution, we employ 
BAM~\cite{Bruegmann:2006ulg,Thierfelder:2011yi,Dietrich:2015iva,Bernuzzi:2016pie}. 
BAM is based on the method-of-lines using a fourth-order explicit Runge-Kutta type integrator, Cartesian grids and finite differencing. 
BAM's grid is based on a hierarchy of cell-centered nested Cartesian boxes consisting of $L$ refinement levels. 
Each level's resolution increases by a factor of two leading to a resolution of $h_l= 2^{-l} h_0$, while $l=0$ is the coarsest level. 
The outer levels remain fixed and employ $n^3/2$ grid points, where the factor $1/2$ arrives for the employed bitant symmetry. 
The inner levels employ $n_{\rm mv}$ points per direction and move with the star's center. 
For the time stepping of the refinement level, we employ the Berger-Oliger scheme~\cite{Berger:1984zza}.
The equations of general relativistic hydrodynamics are solved with high-resolution-shock-capturing schemes based on primitive reconstruction and the local Lax-Friedrich central scheme for the numerical fluxes; see Refs.~\cite{Thierfelder:2011yi,Bernuzzi:2012ci} for more details.
The exact grid configurations are given explicitly in Tab.~\ref{tab:BAM_grid}.

\begin{figure*}[t]
\centering
        \hspace{0.3cm}\includegraphics[width=1.92\columnwidth]{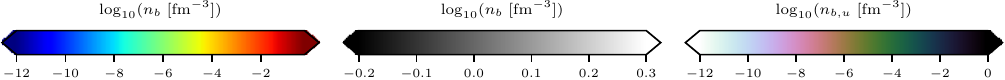}\\
        \vspace{0.3cm}
  \includegraphics[width=0.48\columnwidth]{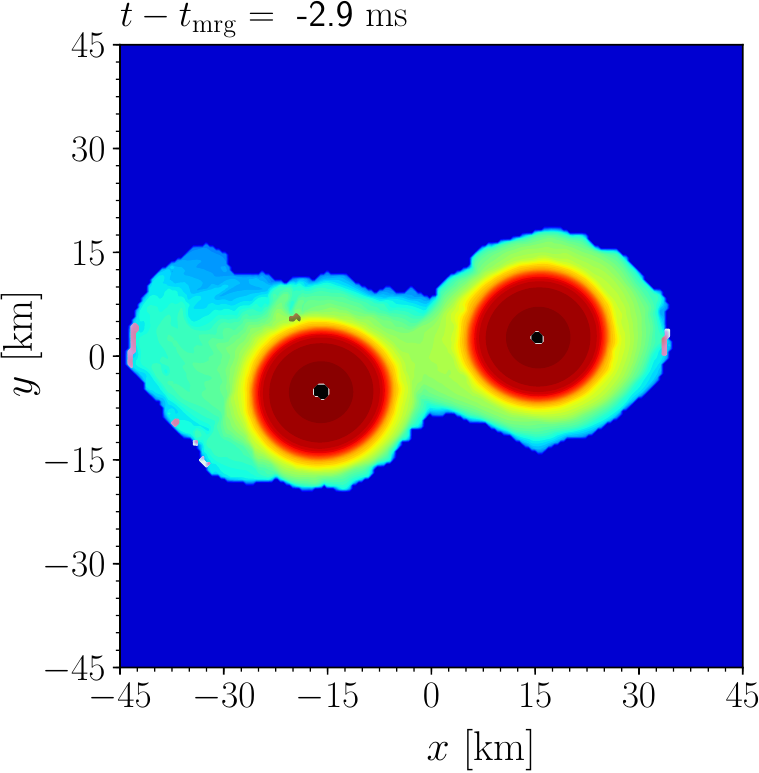} 
    \includegraphics[width=0.48\columnwidth]{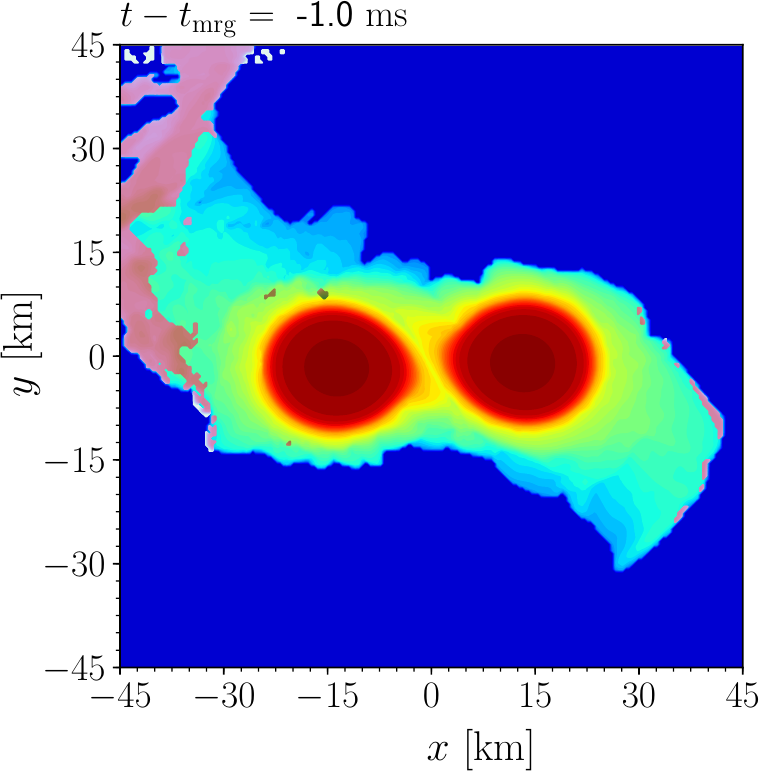} 
      \includegraphics[width=0.48\columnwidth]{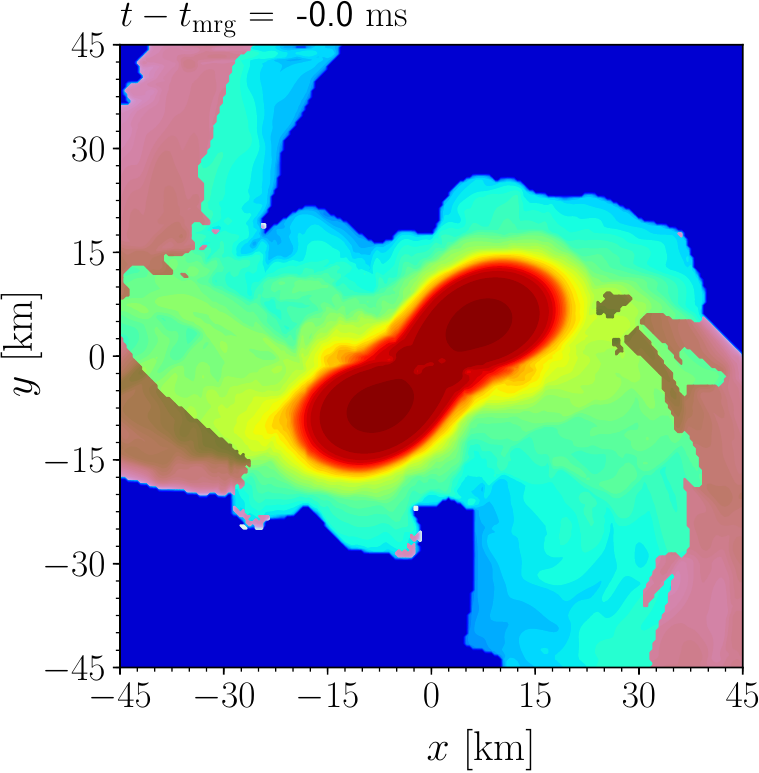} 
        \includegraphics[width=0.48\columnwidth]{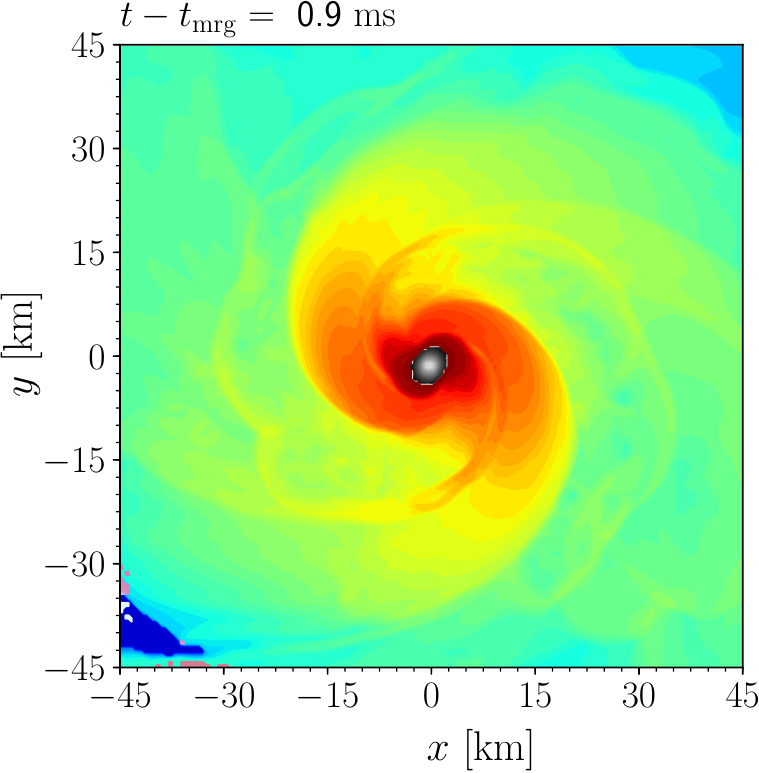} \\
	\hspace{0.05cm}\includegraphics[width=0.48\columnwidth, height=0.4\columnwidth]{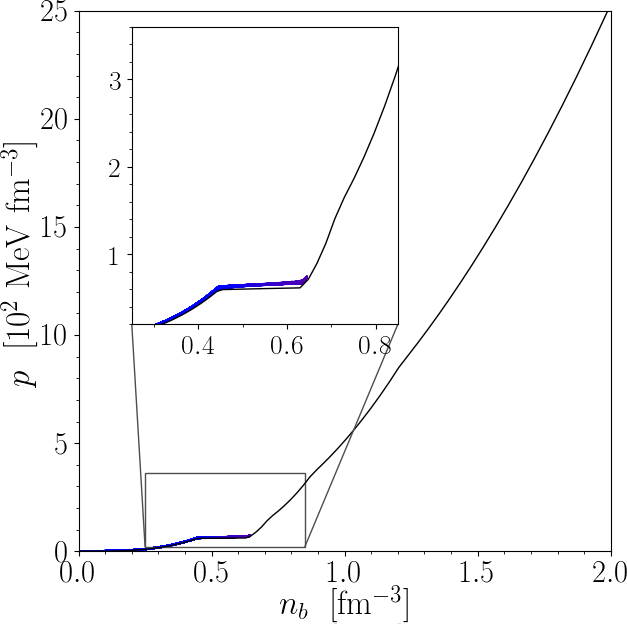}
 \includegraphics[width=0.48\columnwidth, height=0.4\columnwidth]{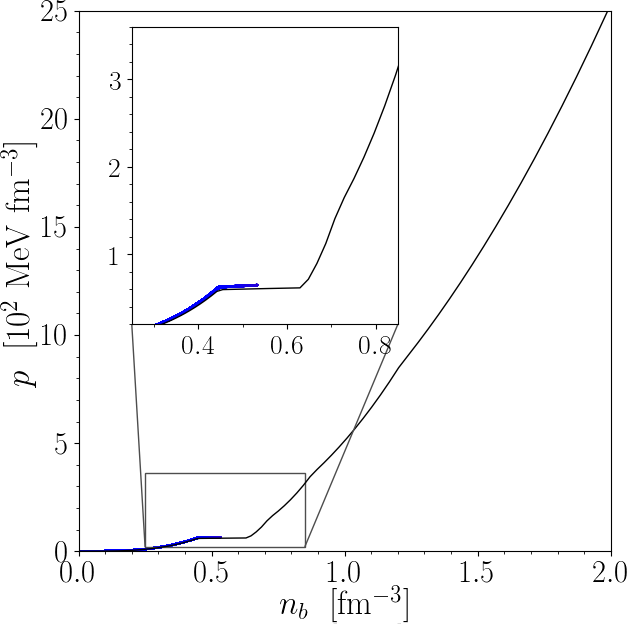}
      \includegraphics[width=0.48\columnwidth,height=0.4\columnwidth]{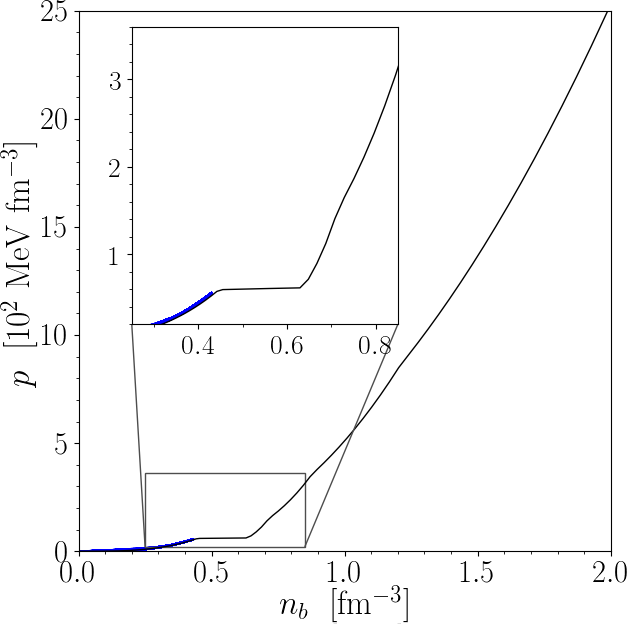}
        \includegraphics[width=0.48\columnwidth,height=0.4\columnwidth]{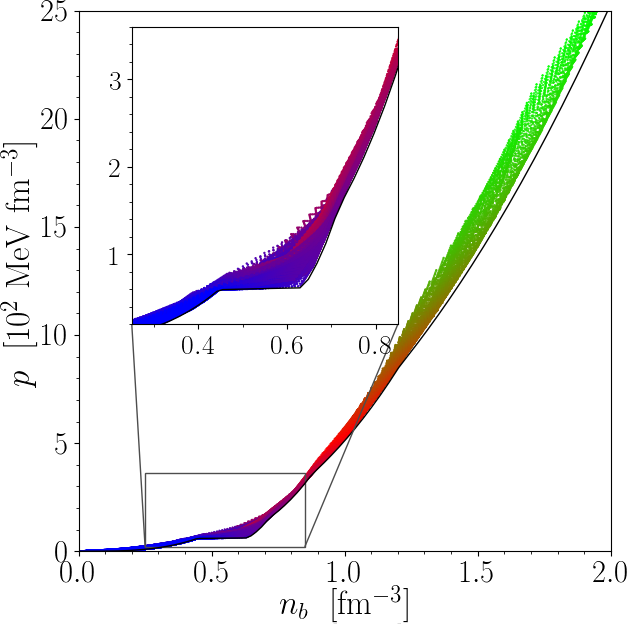}\\
                \hspace{0.3cm}\includegraphics[width=1.92\columnwidth]{colorbar_epsilon.pdf}
\caption{Snapshots of the baryon number density in the equatorial plane (upper panels) and the respective baryon number density-pressure phase diagram (lower panels) for the PT1 case employing the highest resolution. 
         We represent the hadronic matter density in the upper left colorbar, the quark matter density in upper middle colorbar and the ejecta density in the upper right colorbar.
         The lower colorbar refers to the specific internal energy. 
         \textit{Upper panels, first column}: at this instance, the inner core contains quarks at $\log_{10}(n_b~[{\rm fm^{-3}}]) \sim -0.1$. 
         \textit{Upper panels, second column}: the stars are tidally deformed, decreasing the core densities. 
         The inner cores no longer contain quarks, since they reach maximum densities of $\log_{10}(n_b~[{\rm fm^{-3}}]) \sim -0.3$, below the threshold value at which quarks are present. 
         \textit{Upper panels, third column}: moment of merger, where we remark the complete absence of quarks in the coalescing material. 
         \textit{Upper panels, fourth column}: in less than $1~{\rm ms}$ after the merger, a sizeable quark-matter core is formed, which then quickly collapses to a BH. 
         \textit{Lower panels}: The zero-temperature PT1 EOS is depicted (black line) along with the color-coded specific internal energies for the respective snapshots; from this plot it is clear that at late stages of the inspiral, the quarks in the core of both stars vanish, marking ``the reverse phase transition''.} 
\label{fig:2D_PT01}
\end{figure*}

\section{Reverse Phase Transition}
\label{sec:results}

\textbf{Dynamical evolution:}
We start our discussion by presenting the evolution of the baryon number density for the noPT1 scenario inside the orbital plane (upper panels of Fig.~\ref{fig:2D_noPT1}). 
In the lower panels we show the corresponding density-pressure region that is covered in our simulation using the same data as shown in the respective upper panels. 
We mark the zero-temperature EOS as a solid black line. 
We find that during the evolution, despite the fact that the system is in principle symmetric, slight differences, that are present during the construction of the initial data, lead to an evolution that shows a different behavior for both stars, with one star being more deformed and developing a low-density tail. 
As expected, and shown in the bottom panels, temperature effects are not significant before the merger of the two stars, and the total pressure remains close to the zero-temperature EOS.
Only around the time of merger (third column), the increasing temperature leads to a pressure that lies above the zero-temperature curve. 
This becomes even more pronounced in the post-merger phase (fourth column).
In this phase, the densities and pressures increase significantly, while the range of pressures and densities stays almost constant during the inspiral. 
The increase of the central density leads to a prompt gravitational collapse of the remnant and to black hole formation.
Indeed, during the collapse, the simulation reaches maximum baryon number densities of $n_b \geq 5 ~{\rm fm^{-3}}$.
As visible in the last column of Fig.~\ref{fig:2D_noPT1}, most of the mass ejection happens during the early post-merger phase. 

\begin{figure}[htpb]
\centering
  \includegraphics[width=0.85\columnwidth]{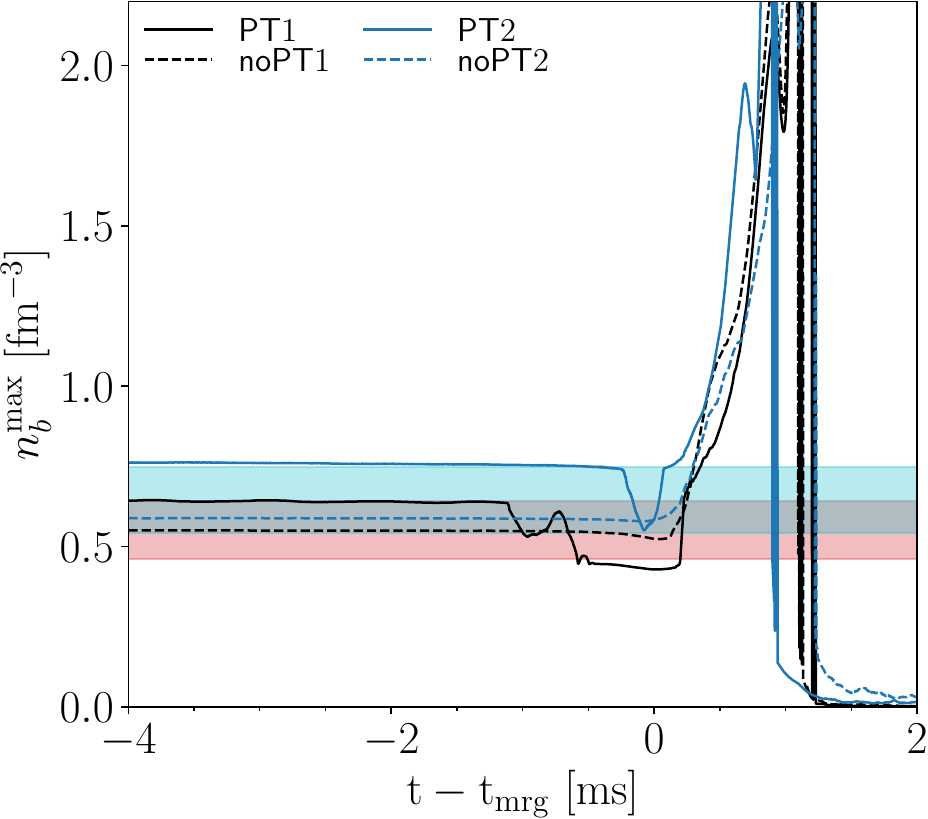}\\
\caption{Maximum baryon number density evolution of all simulations for the highest resolution R4. The red filled region corresponds to the density interval spanning the phase transition for the PT1 EoS. The blue filled region spans the range of densities of the phase transition for the PT2 EoS. Here we note that all runs exhibit almost no density oscillations before the merger. The thick lines representing PT1 (black) and PT2 (blue) stay above their respective onset densities during the inspiral, which indicates the presence of quarks. Around 1 ms before the merger the PT1 and PT2 maximum densities drop noticeably to values below the onset of the phase transition, evidencing the reverse phase transition. Comparatively, noPT1 and noPT2 (dashed lines) experience a smaller decrease of maximum density. Finally, all setups show the characteristic increase of maximum density within 1 ms after the merger, signaling the prompt BH formation.}
\label{fig:rho_max}
\end{figure}

For the PT1 setup, where $q=1$, the maximum density reached inside the two individual stars in the binary lies above the onset density of the phase transition, see Fig.~\ref{fig:EOS_overview}.
Hence, during the inspiral, both stars contain an exotic-matter core; cf.~Fig.~\ref{fig:2D_PT01}.
In the following, we will assume that this core is made from quark matter.
When the stars come closer and tidal interactions become important, the deformation of the star leads to a decrease of the central density . 
While such a decrease is also present for the noPT1 scenario and has also been seen in previous works, e.g.~Fig.~2 of~\cite{Weih:2019xvw}, the presence of a phase transition increases this drop in density noticeably cf.~Fig.~\ref{fig:rho_max}, where the time evolution of the maximum baryon density is shown.
In our case, no quark matter is present shortly before the merger (third column), i.e., the stars undergo a ``reverse phase transition''. 
Once the stars merge, the central densities rapidly increase and the remnant again has a sizeable quark-matter core until the central region quickly collapses to a black hole (not shown in the figure). 
Compared to the PT1 scenario, significantly less mass is ejected at this time; compare color bar of $n_{b,u}$. 

We note that a similar reverse phase transition is also observable in the PT2 setup for which we present the corresponding figures in Appendix~\ref{app:extra}, together with the evolution of noPT2. 
Furthermore, it is worth pointing out that the presence of the reverse phase transition is independent of the employed resolution, i.e., all our setups for various resolutions (Tab.~\ref{tab:BAM_grid}) show similar features. 
We also tested different resolutions of the initial data to verify our findings and found consistent results. 
Overall, this leads to the suggestion that the observed feature is robust and not a numerical artifact. 
Nevertheless, we want to present a few words of caution: we model the microphysical EOS simply through a pressure vs.\ energy density relation, while the thermal contributions to thermodynamical quantities are introduced ad-hoc by an additional ideal-gas contribution. As a consequence, temperature effects due to a change of the degrees of freedom around the reverse phase transition are not taken into account.

As mentioned in Sec.~\ref{sec:intro}, the reverse phase transition described in this section is different in nature with the one presented in \cite{Liebling:2020dhf}, where the change in composition inside the star is encountered in the post-merger phase due to oscillations of the central mass density.

\begin{figure}
\centering
  \centering
  \includegraphics[width=\columnwidth]{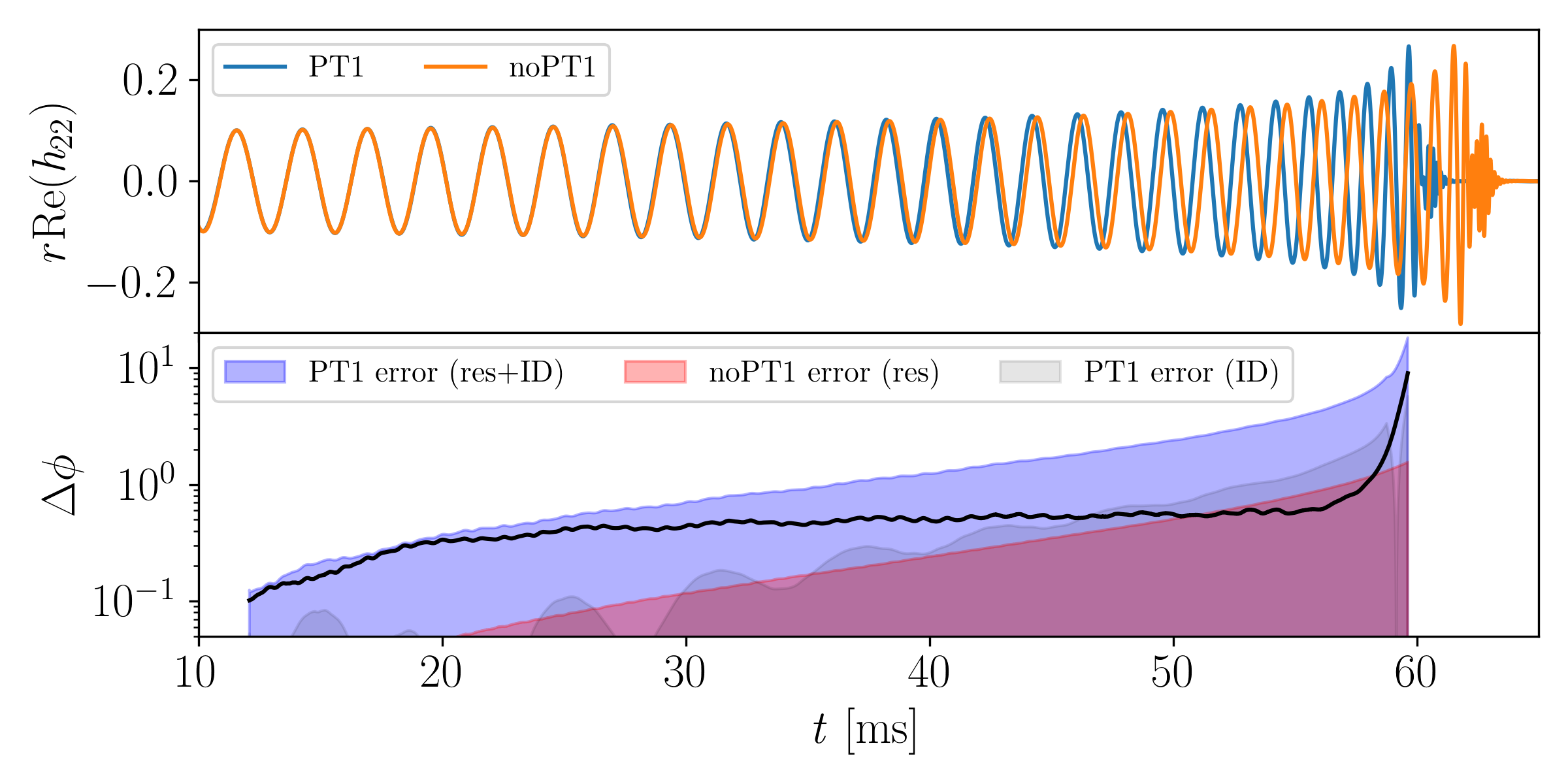}
\caption{Top panel: (2,-2) mode of highest resolution (R4) waveform for the PT1 (blue) and noPT1 (orange) scenarios. 
{Bottom panel:} The black line shows the phase difference between the Richardson extrapolated waveforms for PT1 and noPT1. The red-shaded area shows the error on the GW phase $\phi$ for noPT1 due to the finite resolutions effects computed with Eq. (\ref{eq:delta_res}), while the gray-shaded area shows the error coming from the ID resolution for PT1, Eq. (\ref{eq:delta_ID}). 
Finally the blue-shaded area shows the total error on PT1 computed following Eq. (\ref{eq:delta_tot}).
} 
\label{fig:GW1}
\end{figure}

\textbf{GW emission:}
Given the particular setup of our configurations, in which we picked the same total masses, mass ratios, and tidal deformabilties for PT1/noPT1 and PT2/noPT2, respectively, one would expect that differences in the extracted GW signals, if present, will mainly\footnote{Our setups do not ensure that higher-order tidal effects are identical, which, in principle could also lead to small differences. These are, however, much smaller than the differences that we find in our simulations.} 
arise from the presence/absence of a phase transition. 
A comparison for PT1 and noPT1 is shown in Fig.~\ref{fig:GW1} for the highest resolutions (R4). Overall, we find a visible dephasing between
both setups, but these differences are not significant compared to the large uncertainty of our simulations. The uncertainty itself is constructed using Richardson extrapolation assuming first-order convergence for the PT1 setups and second-order for noPT1 and include an additional uncertainty that is incorporated because of uncertainties in the initial configurations; cf.~Appendix~\ref{app:waves}. 
Overall, the uncertainty is dominated by finite resolution effects for the PT cases. 
Such a behavior was already observed before~\cite{Gieg:2019yzq} and is caused by the non-differentiability of $p(\rho)$ inside the star.
Independent of the large uncertainty and the fact that we can not robustly distinguish the two waveforms, we want to highlight the fact that close to the moment of merger, when the reverse phase transition sets in, no surprising or artificial feature is present in the PT1 case (or similarly the PT2). 
Hence, we suggest that, given the current accuracy of our simulations, we are not able to find a clear imprint of the reverse phase transition on the observable GW signal.

\begin{table}[t]
\caption{Maximum value of ejecta mass $M_{\rm ej}^{\rm max}$ for the two highest resolutions of Tab.~\ref{tab:BAM_grid}. The left column refers to the name of the run and the right columns correspond to the maximum ejecta mass for resolution R4 and R3, respectively.} \label{tab:ejecta}
\begin{tabular}{c|cc}
\toprule
Name & ${}_{\rm R4}M_{\rm ej}^{\max}~[\times 10^{-3}~M_\odot]$ & ${}_{\rm R3}M_{\rm ej}^{\max}~[\times 10^{-3}~M_\odot]$ \\ \hline \hline
PT1 & $0.159$ & $0.158$\\
noPT1 & $1.62$ & $2.22$\\
PT2 & $0.692$ & $0.434$\\
noPT2 & $10.1$ & $12.0$\\
\hline \hline
\end{tabular}
\end{table}

\textbf{Mass ejection:}
In addition to the emission of GWs, BNS mergers emit EM signatures such as the kilonova that is created by the outflowing neutron-rich material released during and after the collision of the two stars. 
To estimate the effect on the EM signal, we present the mass of the ejected material in Tab.~\ref{tab:ejecta} for the two highest resolutions. 
Interestingly, we see that the PT setups eject significantly less (about an order of magnitude) material than noPT setups, despite the almost identical time interval to the gravitational collapse of the remnants and to prompt black hole formation.
Hence, these setups would generally create much dimmer kilonovae. 
For both the PT1 and noPT1 setups, the ejecta are produced within the first millisecond after the merger which suggests that the main ejection mechanism is related to the core bounce.
Hence, we assume that the reverse phase transition has no observable impact on the ejecta.

In fact, looking at baryon densities above 1.5~fm$^{-3}$, presented in the $n_b$-$p$-$\epsilon$ diagrams of Fig.~\ref{fig:2D_noPT1} and Fig.~\ref{fig:2D_PT01} (lower panels, fourth column), we see that the PT1 setup is less populated than the noPT1 setup. 
Also, for higher densities $\approx 1.9/2.0$~fm$^{-3}$, which corresponds to the core of the neutron stars, the noPT1 case reach higher pressures due to a higher thermal contribution. 
This suggests that the noPT1 remnant core exerts larger pressures on the outer layers, which could explain the larger amount of core-bounce ejecta. 

\section{Conclusion}
\label{sec:conclusion}

Binary NS mergers are an exciting class of astrophysical transients that provide us with important information about the behavior of dense matter at supranuclear densities. 
With the help of new numerical-relativity simulations, we have investigated the merger dynamics of systems in which the stars undergo a reverse phase transition before the merger, i.e., the simulated quark core vanishes for the last milliseconds before the collision of the stars. After the collision, a quark core is quickly formed again and, for the particular EOSs that we employed, the remnant collapses promptly to a black hole. 
We verified that the presence of the reverse phase transition happened independent of the simulated setup as long as the onset density was close to the central density of the individual stars, the resolution of the initial data, and the resolution employed during the dynamical evolution. 

Despite the fact that the reverse phase transition is certainly of academic interest, we did not find an imprint on the gravitational-wave signal or on the mass ejection that is noticeable given the numerical precision of our simulations.
However, should future high-precision simulations show that the reverse phase transition is indeed leading to an observable phase difference, additional follow-up studies may be warranted. Due to the prompt BH formation of the remnant, we were not able to verify any imprints of the phase transition (to quark matter at merger) in the post-merger gravitational wave signals, as found in \cite{Most:2018eaw,Bauswein:2018bma,Weih:2019xvw,Blacker:2020nlq,Bauswein:2020ggy,Huang:2022mqp}. Nevertheless, our simulations verified, in agreement with previous works, that the presence of a phase transition can noticeably soften the EOS at merger time~\cite{Most:2018eaw,Bauswein:2018bma,Blacker:2020nlq,Bauswein:2020ggy,Prakash:2021wpz} and can alter the electromagnetic signals from a BNS merger~\cite{Prakash:2021wpz} due to different ejecta properties. Hence, the joined observation of gravitational waves and electromagnetic signals from a binary NS merger will enable a better understanding of the EOS of supranuclear dense matter. 

\acknowledgments{
M.U and H.G. acknowledge FAPESP for financial support under grant number 2019/26287-0.
The work of I.T. was supported by the U.S. Department of Energy, Office of Science, Office of Nuclear Physics, under Contract No.~DE-AC52-06NA25396, by the Laboratory Directed Research and Development program of Los Alamos National Laboratory under Project Nos.\ 20220658ER and 20230315ER, and by the U.S. Department of Energy, Office of Science, Office of Advanced Scientific Computing Research, Scientific Discovery through Advanced Computing (SciDAC) NUCLEI program. S.V.C. was supported by the research environment grant ``Gravitational Radiation and Electromagnetic Astrophysical Transients (GREAT)'' funded by the Swedish Research council (VR) under Grant No. Dnr. 2016-06012.

The simulations were performed on the national supercomputer HPE Apollo Hawk at the High Performance Computing (HPC) Center Stuttgart (HLRS) under the grant number GWanalysis/44189, on the GCS Supercomputer SuperMUC\_NG at the Leibniz Supercomputing Centre (LRZ) [project pn29ba], and on the HPC systems Lise/Emmy of the North German Supercomputing Alliance (HLRN) [project bbp00049]. 
}

%%%%%%%%%%%%%%%%%%%%%%%%%%%%%%%%%%%%

\bibliography{nr_pt.bib}

\begin{thebibliography}{67}%
\makeatletter
\providecommand \@ifxundefined [1]{%
 \@ifx{#1\undefined}
}%
\providecommand \@ifnum [1]{%
 \ifnum #1\expandafter \@firstoftwo
 \else \expandafter \@secondoftwo
 \fi
}%
\providecommand \@ifx [1]{%
 \ifx #1\expandafter \@firstoftwo
 \else \expandafter \@secondoftwo
 \fi
}%
\providecommand \natexlab [1]{#1}%
\providecommand \enquote  [1]{``#1''}%
\providecommand \bibnamefont  [1]{#1}%
\providecommand \bibfnamefont [1]{#1}%
\providecommand \citenamefont [1]{#1}%
\providecommand \href@noop [0]{\@secondoftwo}%
\providecommand \href [0]{\begingroup \@sanitize@url \@href}%
\providecommand \@href[1]{\@@startlink{#1}\@@href}%
\providecommand \@@href[1]{\endgroup#1\@@endlink}%
\providecommand \@sanitize@url [0]{\catcode `\\12\catcode `\$12\catcode
  `\&12\catcode `\#12\catcode `\^12\catcode `\_12\catcode `\%12\relax}%
\providecommand \@@startlink[1]{}%
\providecommand \@@endlink[0]{}%
\providecommand \url  [0]{\begingroup\@sanitize@url \@url }%
\providecommand \@url [1]{\endgroup\@href {#1}{\urlprefix }}%
\providecommand \urlprefix  [0]{URL }%
\providecommand \Eprint [0]{\href }%
\providecommand \doibase [0]{https://doi.org/}%
\providecommand \selectlanguage [0]{\@gobble}%
\providecommand \bibinfo  [0]{\@secondoftwo}%
\providecommand \bibfield  [0]{\@secondoftwo}%
\providecommand \translation [1]{[#1]}%
\providecommand \BibitemOpen [0]{}%
\providecommand \bibitemStop [0]{}%
\providecommand \bibitemNoStop [0]{.\EOS\space}%
\providecommand \EOS [0]{\spacefactor3000\relax}%
\providecommand \BibitemShut  [1]{\csname bibitem#1\endcsname}%
\let\auto@bib@innerbib\@empty
%</preamble>
\bibitem [{\citenamefont {Huth}\ \emph {et~al.}(2022)\citenamefont {Huth} \emph
  {et~al.}}]{Huth:2021bsp}%
  \BibitemOpen
  \bibfield  {author} {\bibinfo {author} {\bibfnamefont {S.}~\bibnamefont
  {Huth}} \emph {et~al.},\ }\bibfield  {title} {\bibinfo {title} {{Constraining
  Neutron-Star Matter with Microscopic and Macroscopic Collisions}},\ }\href
  {https://doi.org/10.1038/s41586-022-04750-w} {\bibfield  {journal} {\bibinfo
  {journal} {Nature}\ }\textbf {\bibinfo {volume} {606}},\ \bibinfo {pages}
  {276} (\bibinfo {year} {2022})},\ \Eprint {https://arxiv.org/abs/2107.06229}
  {arXiv:2107.06229 [nucl-th]} \BibitemShut {NoStop}%
\bibitem [{\citenamefont {Abbott}\ \emph
  {et~al.}(2017{\natexlab{a}})\citenamefont {Abbott} \emph
  {et~al.}}]{LIGOScientific:2017vwq}%
  \BibitemOpen
  \bibfield  {author} {\bibinfo {author} {\bibfnamefont {B.~P.}\ \bibnamefont
  {Abbott}} \emph {et~al.} (\bibinfo {collaboration} {LIGO Scientific,
  Virgo}),\ }\bibfield  {title} {\bibinfo {title} {{GW170817: Observation of
  Gravitational Waves from a Binary Neutron Star Inspiral}},\ }\href
  {https://doi.org/10.1103/PhysRevLett.119.161101} {\bibfield  {journal}
  {\bibinfo  {journal} {Phys. Rev. Lett.}\ }\textbf {\bibinfo {volume} {119}},\
  \bibinfo {pages} {161101} (\bibinfo {year} {2017}{\natexlab{a}})},\ \Eprint
  {https://arxiv.org/abs/1710.05832} {arXiv:1710.05832 [gr-qc]} \BibitemShut
  {NoStop}%
\bibitem [{\citenamefont {Abbott}\ \emph
  {et~al.}(2017{\natexlab{b}})\citenamefont {Abbott} \emph
  {et~al.}}]{LIGOScientific:2017zic}%
  \BibitemOpen
  \bibfield  {author} {\bibinfo {author} {\bibfnamefont {B.~P.}\ \bibnamefont
  {Abbott}} \emph {et~al.} (\bibinfo {collaboration} {LIGO Scientific, Virgo,
  Fermi-GBM, INTEGRAL}),\ }\bibfield  {title} {\bibinfo {title} {{Gravitational
  Waves and Gamma-rays from a Binary Neutron Star Merger: GW170817 and GRB
  170817A}},\ }\href {https://doi.org/10.3847/2041-8213/aa920c} {\bibfield
  {journal} {\bibinfo  {journal} {Astrophys. J. Lett.}\ }\textbf {\bibinfo
  {volume} {848}},\ \bibinfo {pages} {L13} (\bibinfo {year}
  {2017}{\natexlab{b}})},\ \Eprint {https://arxiv.org/abs/1710.05834}
  {arXiv:1710.05834 [astro-ph.HE]} \BibitemShut {NoStop}%
\bibitem [{\citenamefont {Dietrich}\ \emph {et~al.}(2020)\citenamefont
  {Dietrich}, \citenamefont {Coughlin}, \citenamefont {Pang}, \citenamefont
  {Bulla}, \citenamefont {Heinzel}, \citenamefont {Issa}, \citenamefont
  {Tews},\ and\ \citenamefont {Antier}}]{Dietrich:2020efo}%
  \BibitemOpen
  \bibfield  {author} {\bibinfo {author} {\bibfnamefont {T.}~\bibnamefont
  {Dietrich}}, \bibinfo {author} {\bibfnamefont {M.~W.}\ \bibnamefont
  {Coughlin}}, \bibinfo {author} {\bibfnamefont {P.~T.~H.}\ \bibnamefont
  {Pang}}, \bibinfo {author} {\bibfnamefont {M.}~\bibnamefont {Bulla}},
  \bibinfo {author} {\bibfnamefont {J.}~\bibnamefont {Heinzel}}, \bibinfo
  {author} {\bibfnamefont {L.}~\bibnamefont {Issa}}, \bibinfo {author}
  {\bibfnamefont {I.}~\bibnamefont {Tews}},\ and\ \bibinfo {author}
  {\bibfnamefont {S.}~\bibnamefont {Antier}},\ }\bibfield  {title} {\bibinfo
  {title} {{Multimessenger constraints on the neutron-star equation of state
  and the Hubble constant}},\ }\href {https://doi.org/10.1126/science.abb4317}
  {\bibfield  {journal} {\bibinfo  {journal} {Science}\ }\textbf {\bibinfo
  {volume} {370}},\ \bibinfo {pages} {1450} (\bibinfo {year} {2020})},\ \Eprint
  {https://arxiv.org/abs/2002.11355} {arXiv:2002.11355 [astro-ph.HE]}
  \BibitemShut {NoStop}%
\bibitem [{\citenamefont {Baiotti}\ and\ \citenamefont
  {Rezzolla}(2017)}]{Baiotti:2016qnr}%
  \BibitemOpen
  \bibfield  {author} {\bibinfo {author} {\bibfnamefont {L.}~\bibnamefont
  {Baiotti}}\ and\ \bibinfo {author} {\bibfnamefont {L.}~\bibnamefont
  {Rezzolla}},\ }\bibfield  {title} {\bibinfo {title} {{Binary neutron star
  mergers: a review of Einstein\textquoteright{}s richest laboratory}},\ }\href
  {https://doi.org/10.1088/1361-6633/aa67bb} {\bibfield  {journal} {\bibinfo
  {journal} {Rept. Prog. Phys.}\ }\textbf {\bibinfo {volume} {80}},\ \bibinfo
  {pages} {096901} (\bibinfo {year} {2017})},\ \Eprint
  {https://arxiv.org/abs/1607.03540} {arXiv:1607.03540 [gr-qc]} \BibitemShut
  {NoStop}%
\bibitem [{\citenamefont {Baiotti}(2019)}]{Baiotti:2019sew}%
  \BibitemOpen
  \bibfield  {author} {\bibinfo {author} {\bibfnamefont {L.}~\bibnamefont
  {Baiotti}},\ }\bibfield  {title} {\bibinfo {title} {{Gravitational waves from
  neutron star mergers and their relation to the nuclear equation of state}},\
  }\href {https://doi.org/10.1016/j.ppnp.2019.103714} {\bibfield  {journal}
  {\bibinfo  {journal} {Prog. Part. Nucl. Phys.}\ }\textbf {\bibinfo {volume}
  {109}},\ \bibinfo {pages} {103714} (\bibinfo {year} {2019})},\ \Eprint
  {https://arxiv.org/abs/1907.08534} {arXiv:1907.08534 [astro-ph.HE]}
  \BibitemShut {NoStop}%
\bibitem [{\citenamefont {Dietrich}\ \emph {et~al.}(2021)\citenamefont
  {Dietrich}, \citenamefont {Hinderer},\ and\ \citenamefont
  {Samajdar}}]{Dietrich:2020eud}%
  \BibitemOpen
  \bibfield  {author} {\bibinfo {author} {\bibfnamefont {T.}~\bibnamefont
  {Dietrich}}, \bibinfo {author} {\bibfnamefont {T.}~\bibnamefont {Hinderer}},\
  and\ \bibinfo {author} {\bibfnamefont {A.}~\bibnamefont {Samajdar}},\
  }\bibfield  {title} {\bibinfo {title} {{Interpreting Binary Neutron Star
  Mergers: Describing the Binary Neutron Star Dynamics, Modelling Gravitational
  Waveforms, and Analyzing Detections}},\ }\href
  {https://doi.org/10.1007/s10714-020-02751-6} {\bibfield  {journal} {\bibinfo
  {journal} {Gen. Rel. Grav.}\ }\textbf {\bibinfo {volume} {53}},\ \bibinfo
  {pages} {27} (\bibinfo {year} {2021})},\ \Eprint
  {https://arxiv.org/abs/2004.02527} {arXiv:2004.02527 [gr-qc]} \BibitemShut
  {NoStop}%
\bibitem [{\citenamefont {Radice}\ \emph {et~al.}(2020)\citenamefont {Radice},
  \citenamefont {Bernuzzi},\ and\ \citenamefont {Perego}}]{Radice:2020ddv}%
  \BibitemOpen
  \bibfield  {author} {\bibinfo {author} {\bibfnamefont {D.}~\bibnamefont
  {Radice}}, \bibinfo {author} {\bibfnamefont {S.}~\bibnamefont {Bernuzzi}},\
  and\ \bibinfo {author} {\bibfnamefont {A.}~\bibnamefont {Perego}},\
  }\bibfield  {title} {\bibinfo {title} {{The Dynamics of Binary Neutron Star
  Mergers and GW170817}},\ }\href
  {https://doi.org/10.1146/annurev-nucl-013120-114541} {\bibfield  {journal}
  {\bibinfo  {journal} {Ann. Rev. Nucl. Part. Sci.}\ }\textbf {\bibinfo
  {volume} {70}},\ \bibinfo {pages} {95} (\bibinfo {year} {2020})},\ \Eprint
  {https://arxiv.org/abs/2002.03863} {arXiv:2002.03863 [astro-ph.HE]}
  \BibitemShut {NoStop}%
\bibitem [{\citenamefont {Radice}\ \emph {et~al.}(2014)\citenamefont {Radice},
  \citenamefont {Rezzolla},\ and\ \citenamefont {Galeazzi}}]{Radice:2013hxh}%
  \BibitemOpen
  \bibfield  {author} {\bibinfo {author} {\bibfnamefont {D.}~\bibnamefont
  {Radice}}, \bibinfo {author} {\bibfnamefont {L.}~\bibnamefont {Rezzolla}},\
  and\ \bibinfo {author} {\bibfnamefont {F.}~\bibnamefont {Galeazzi}},\
  }\bibfield  {title} {\bibinfo {title} {{Beyond second-order convergence in
  simulations of binary neutron stars in full general-relativity}},\ }\href
  {https://doi.org/10.1093/mnrasl/slt137} {\bibfield  {journal} {\bibinfo
  {journal} {Mon. Not. Roy. Astron. Soc.}\ }\textbf {\bibinfo {volume} {437}},\
  \bibinfo {pages} {L46} (\bibinfo {year} {2014})},\ \Eprint
  {https://arxiv.org/abs/1306.6052} {arXiv:1306.6052 [gr-qc]} \BibitemShut
  {NoStop}%
\bibitem [{\citenamefont {Hotokezaka}\ \emph {et~al.}(2015)\citenamefont
  {Hotokezaka}, \citenamefont {Kyutoku}, \citenamefont {Okawa},\ and\
  \citenamefont {Shibata}}]{Hotokezaka:2015xka}%
  \BibitemOpen
  \bibfield  {author} {\bibinfo {author} {\bibfnamefont {K.}~\bibnamefont
  {Hotokezaka}}, \bibinfo {author} {\bibfnamefont {K.}~\bibnamefont {Kyutoku}},
  \bibinfo {author} {\bibfnamefont {H.}~\bibnamefont {Okawa}},\ and\ \bibinfo
  {author} {\bibfnamefont {M.}~\bibnamefont {Shibata}},\ }\bibfield  {title}
  {\bibinfo {title} {{Exploring tidal effects of coalescing binary neutron
  stars in numerical relativity. II. Long-term simulations}},\ }\href
  {https://doi.org/10.1103/PhysRevD.91.064060} {\bibfield  {journal} {\bibinfo
  {journal} {Phys. Rev. D}\ }\textbf {\bibinfo {volume} {91}},\ \bibinfo
  {pages} {064060} (\bibinfo {year} {2015})},\ \Eprint
  {https://arxiv.org/abs/1502.03457} {arXiv:1502.03457 [gr-qc]} \BibitemShut
  {NoStop}%
\bibitem [{\citenamefont {Kiuchi}\ \emph {et~al.}(2017)\citenamefont {Kiuchi},
  \citenamefont {Kawaguchi}, \citenamefont {Kyutoku}, \citenamefont
  {Sekiguchi}, \citenamefont {Shibata},\ and\ \citenamefont
  {Taniguchi}}]{Kiuchi:2017pte}%
  \BibitemOpen
  \bibfield  {author} {\bibinfo {author} {\bibfnamefont {K.}~\bibnamefont
  {Kiuchi}}, \bibinfo {author} {\bibfnamefont {K.}~\bibnamefont {Kawaguchi}},
  \bibinfo {author} {\bibfnamefont {K.}~\bibnamefont {Kyutoku}}, \bibinfo
  {author} {\bibfnamefont {Y.}~\bibnamefont {Sekiguchi}}, \bibinfo {author}
  {\bibfnamefont {M.}~\bibnamefont {Shibata}},\ and\ \bibinfo {author}
  {\bibfnamefont {K.}~\bibnamefont {Taniguchi}},\ }\bibfield  {title} {\bibinfo
  {title} {{Sub-radian-accuracy gravitational waveforms of coalescing binary
  neutron stars in numerical relativity}},\ }\href
  {https://doi.org/10.1103/PhysRevD.96.084060} {\bibfield  {journal} {\bibinfo
  {journal} {Phys. Rev. D}\ }\textbf {\bibinfo {volume} {96}},\ \bibinfo
  {pages} {084060} (\bibinfo {year} {2017})},\ \Eprint
  {https://arxiv.org/abs/1708.08926} {arXiv:1708.08926 [astro-ph.HE]}
  \BibitemShut {NoStop}%
\bibitem [{\citenamefont {Bernuzzi}\ and\ \citenamefont
  {Dietrich}(2016)}]{Bernuzzi:2016pie}%
  \BibitemOpen
  \bibfield  {author} {\bibinfo {author} {\bibfnamefont {S.}~\bibnamefont
  {Bernuzzi}}\ and\ \bibinfo {author} {\bibfnamefont {T.}~\bibnamefont
  {Dietrich}},\ }\bibfield  {title} {\bibinfo {title} {{Gravitational waveforms
  from binary neutron star mergers with high-order
  weighted-essentially-nonoscillatory schemes in numerical relativity}},\
  }\href {https://doi.org/10.1103/PhysRevD.94.064062} {\bibfield  {journal}
  {\bibinfo  {journal} {Phys. Rev. D}\ }\textbf {\bibinfo {volume} {94}},\
  \bibinfo {pages} {064062} (\bibinfo {year} {2016})},\ \Eprint
  {https://arxiv.org/abs/1604.07999} {arXiv:1604.07999 [gr-qc]} \BibitemShut
  {NoStop}%
\bibitem [{\citenamefont {Anderson}\ \emph {et~al.}(2008)\citenamefont
  {Anderson}, \citenamefont {Hirschmann}, \citenamefont {Lehner}, \citenamefont
  {Liebling}, \citenamefont {Motl}, \citenamefont {Neilsen}, \citenamefont
  {Palenzuela},\ and\ \citenamefont {Tohline}}]{Anderson:2008zp}%
  \BibitemOpen
  \bibfield  {author} {\bibinfo {author} {\bibfnamefont {M.}~\bibnamefont
  {Anderson}}, \bibinfo {author} {\bibfnamefont {E.~W.}\ \bibnamefont
  {Hirschmann}}, \bibinfo {author} {\bibfnamefont {L.}~\bibnamefont {Lehner}},
  \bibinfo {author} {\bibfnamefont {S.~L.}\ \bibnamefont {Liebling}}, \bibinfo
  {author} {\bibfnamefont {P.~M.}\ \bibnamefont {Motl}}, \bibinfo {author}
  {\bibfnamefont {D.}~\bibnamefont {Neilsen}}, \bibinfo {author} {\bibfnamefont
  {C.}~\bibnamefont {Palenzuela}},\ and\ \bibinfo {author} {\bibfnamefont
  {J.~E.}\ \bibnamefont {Tohline}},\ }\bibfield  {title} {\bibinfo {title}
  {{Magnetized Neutron Star Mergers and Gravitational Wave Signals}},\ }\href
  {https://doi.org/10.1103/PhysRevLett.100.191101} {\bibfield  {journal}
  {\bibinfo  {journal} {Phys. Rev. Lett.}\ }\textbf {\bibinfo {volume} {100}},\
  \bibinfo {pages} {191101} (\bibinfo {year} {2008})},\ \Eprint
  {https://arxiv.org/abs/0801.4387} {arXiv:0801.4387 [gr-qc]} \BibitemShut
  {NoStop}%
\bibitem [{\citenamefont {Giacomazzo}\ \emph {et~al.}(2011)\citenamefont
  {Giacomazzo}, \citenamefont {Rezzolla},\ and\ \citenamefont
  {Baiotti}}]{Giacomazzo:2010bx}%
  \BibitemOpen
  \bibfield  {author} {\bibinfo {author} {\bibfnamefont {B.}~\bibnamefont
  {Giacomazzo}}, \bibinfo {author} {\bibfnamefont {L.}~\bibnamefont
  {Rezzolla}},\ and\ \bibinfo {author} {\bibfnamefont {L.}~\bibnamefont
  {Baiotti}},\ }\bibfield  {title} {\bibinfo {title} {{Accurate evolutions of
  inspiralling and magnetized neutron-stars: Equal-mass binaries}},\ }\href
  {https://doi.org/10.1103/PhysRevD.83.044014} {\bibfield  {journal} {\bibinfo
  {journal} {Phys. Rev. D}\ }\textbf {\bibinfo {volume} {83}},\ \bibinfo
  {pages} {044014} (\bibinfo {year} {2011})},\ \Eprint
  {https://arxiv.org/abs/1009.2468} {arXiv:1009.2468 [gr-qc]} \BibitemShut
  {NoStop}%
\bibitem [{\citenamefont {Rezzolla}\ \emph {et~al.}(2011)\citenamefont
  {Rezzolla}, \citenamefont {Giacomazzo}, \citenamefont {Baiotti},
  \citenamefont {Granot}, \citenamefont {Kouveliotou},\ and\ \citenamefont
  {Aloy}}]{Rezzolla:2011da}%
  \BibitemOpen
  \bibfield  {author} {\bibinfo {author} {\bibfnamefont {L.}~\bibnamefont
  {Rezzolla}}, \bibinfo {author} {\bibfnamefont {B.}~\bibnamefont
  {Giacomazzo}}, \bibinfo {author} {\bibfnamefont {L.}~\bibnamefont {Baiotti}},
  \bibinfo {author} {\bibfnamefont {J.}~\bibnamefont {Granot}}, \bibinfo
  {author} {\bibfnamefont {C.}~\bibnamefont {Kouveliotou}},\ and\ \bibinfo
  {author} {\bibfnamefont {M.~A.}\ \bibnamefont {Aloy}},\ }\bibfield  {title}
  {\bibinfo {title} {{The missing link: Merging neutron stars naturally produce
  jet-like structures and can power short Gamma-Ray Bursts}},\ }\href
  {https://doi.org/10.1088/2041-8205/732/1/L6} {\bibfield  {journal} {\bibinfo
  {journal} {Astrophys. J. Lett.}\ }\textbf {\bibinfo {volume} {732}},\
  \bibinfo {pages} {L6} (\bibinfo {year} {2011})},\ \Eprint
  {https://arxiv.org/abs/1101.4298} {arXiv:1101.4298 [astro-ph.HE]}
  \BibitemShut {NoStop}%
\bibitem [{\citenamefont {Kiuchi}\ \emph {et~al.}(2014)\citenamefont {Kiuchi},
  \citenamefont {Kyutoku}, \citenamefont {Sekiguchi}, \citenamefont {Shibata},\
  and\ \citenamefont {Wada}}]{Kiuchi:2014hja}%
  \BibitemOpen
  \bibfield  {author} {\bibinfo {author} {\bibfnamefont {K.}~\bibnamefont
  {Kiuchi}}, \bibinfo {author} {\bibfnamefont {K.}~\bibnamefont {Kyutoku}},
  \bibinfo {author} {\bibfnamefont {Y.}~\bibnamefont {Sekiguchi}}, \bibinfo
  {author} {\bibfnamefont {M.}~\bibnamefont {Shibata}},\ and\ \bibinfo {author}
  {\bibfnamefont {T.}~\bibnamefont {Wada}},\ }\bibfield  {title} {\bibinfo
  {title} {{High resolution numerical-relativity simulations for the merger of
  binary magnetized neutron stars}},\ }\href
  {https://doi.org/10.1103/PhysRevD.90.041502} {\bibfield  {journal} {\bibinfo
  {journal} {Phys. Rev. D}\ }\textbf {\bibinfo {volume} {90}},\ \bibinfo
  {pages} {041502} (\bibinfo {year} {2014})},\ \Eprint
  {https://arxiv.org/abs/1407.2660} {arXiv:1407.2660 [astro-ph.HE]}
  \BibitemShut {NoStop}%
\bibitem [{\citenamefont {Palenzuela}\ \emph {et~al.}(2015)\citenamefont
  {Palenzuela}, \citenamefont {Liebling}, \citenamefont {Neilsen},
  \citenamefont {Lehner}, \citenamefont {Caballero}, \citenamefont {O'Connor},\
  and\ \citenamefont {Anderson}}]{Palenzuela:2015dqa}%
  \BibitemOpen
  \bibfield  {author} {\bibinfo {author} {\bibfnamefont {C.}~\bibnamefont
  {Palenzuela}}, \bibinfo {author} {\bibfnamefont {S.~L.}\ \bibnamefont
  {Liebling}}, \bibinfo {author} {\bibfnamefont {D.}~\bibnamefont {Neilsen}},
  \bibinfo {author} {\bibfnamefont {L.}~\bibnamefont {Lehner}}, \bibinfo
  {author} {\bibfnamefont {O.~L.}\ \bibnamefont {Caballero}}, \bibinfo {author}
  {\bibfnamefont {E.}~\bibnamefont {O'Connor}},\ and\ \bibinfo {author}
  {\bibfnamefont {M.}~\bibnamefont {Anderson}},\ }\bibfield  {title} {\bibinfo
  {title} {{Effects of the microphysical Equation of State in the mergers of
  magnetized Neutron Stars With Neutrino Cooling}},\ }\href
  {https://doi.org/10.1103/PhysRevD.92.044045} {\bibfield  {journal} {\bibinfo
  {journal} {Phys. Rev. D}\ }\textbf {\bibinfo {volume} {92}},\ \bibinfo
  {pages} {044045} (\bibinfo {year} {2015})},\ \Eprint
  {https://arxiv.org/abs/1505.01607} {arXiv:1505.01607 [gr-qc]} \BibitemShut
  {NoStop}%
\bibitem [{\citenamefont {Dionysopoulou}\ \emph {et~al.}(2015)\citenamefont
  {Dionysopoulou}, \citenamefont {Alic},\ and\ \citenamefont
  {Rezzolla}}]{Dionysopoulou:2015tda}%
  \BibitemOpen
  \bibfield  {author} {\bibinfo {author} {\bibfnamefont {K.}~\bibnamefont
  {Dionysopoulou}}, \bibinfo {author} {\bibfnamefont {D.}~\bibnamefont
  {Alic}},\ and\ \bibinfo {author} {\bibfnamefont {L.}~\bibnamefont
  {Rezzolla}},\ }\bibfield  {title} {\bibinfo {title} {{General-relativistic
  resistive-magnetohydrodynamic simulations of binary neutron stars}},\ }\href
  {https://doi.org/10.1103/PhysRevD.92.084064} {\bibfield  {journal} {\bibinfo
  {journal} {Phys. Rev. D}\ }\textbf {\bibinfo {volume} {92}},\ \bibinfo
  {pages} {084064} (\bibinfo {year} {2015})},\ \Eprint
  {https://arxiv.org/abs/1502.02021} {arXiv:1502.02021 [gr-qc]} \BibitemShut
  {NoStop}%
\bibitem [{\citenamefont {Ruiz}\ \emph {et~al.}(2018)\citenamefont {Ruiz},
  \citenamefont {Shapiro},\ and\ \citenamefont {Tsokaros}}]{Ruiz:2017due}%
  \BibitemOpen
  \bibfield  {author} {\bibinfo {author} {\bibfnamefont {M.}~\bibnamefont
  {Ruiz}}, \bibinfo {author} {\bibfnamefont {S.~L.}\ \bibnamefont {Shapiro}},\
  and\ \bibinfo {author} {\bibfnamefont {A.}~\bibnamefont {Tsokaros}},\
  }\bibfield  {title} {\bibinfo {title} {{GW170817, General Relativistic
  Magnetohydrodynamic Simulations, and the Neutron Star Maximum Mass}},\ }\href
  {https://doi.org/10.1103/PhysRevD.97.021501} {\bibfield  {journal} {\bibinfo
  {journal} {Phys. Rev. D}\ }\textbf {\bibinfo {volume} {97}},\ \bibinfo
  {pages} {021501} (\bibinfo {year} {2018})},\ \Eprint
  {https://arxiv.org/abs/1711.00473} {arXiv:1711.00473 [astro-ph.HE]}
  \BibitemShut {NoStop}%
\bibitem [{\citenamefont {Kiuchi}\ \emph {et~al.}(2018)\citenamefont {Kiuchi},
  \citenamefont {Kyutoku}, \citenamefont {Sekiguchi},\ and\ \citenamefont
  {Shibata}}]{Kiuchi:2017zzg}%
  \BibitemOpen
  \bibfield  {author} {\bibinfo {author} {\bibfnamefont {K.}~\bibnamefont
  {Kiuchi}}, \bibinfo {author} {\bibfnamefont {K.}~\bibnamefont {Kyutoku}},
  \bibinfo {author} {\bibfnamefont {Y.}~\bibnamefont {Sekiguchi}},\ and\
  \bibinfo {author} {\bibfnamefont {M.}~\bibnamefont {Shibata}},\ }\bibfield
  {title} {\bibinfo {title} {{Global simulations of strongly magnetized remnant
  massive neutron stars formed in binary neutron star mergers}},\ }\href
  {https://doi.org/10.1103/PhysRevD.97.124039} {\bibfield  {journal} {\bibinfo
  {journal} {Phys. Rev. D}\ }\textbf {\bibinfo {volume} {97}},\ \bibinfo
  {pages} {124039} (\bibinfo {year} {2018})},\ \Eprint
  {https://arxiv.org/abs/1710.01311} {arXiv:1710.01311 [astro-ph.HE]}
  \BibitemShut {NoStop}%
\bibitem [{\citenamefont {Sekiguchi}\ \emph {et~al.}(2011)\citenamefont
  {Sekiguchi}, \citenamefont {Kiuchi}, \citenamefont {Kyutoku},\ and\
  \citenamefont {Shibata}}]{Sekiguchi:2011zd}%
  \BibitemOpen
  \bibfield  {author} {\bibinfo {author} {\bibfnamefont {Y.}~\bibnamefont
  {Sekiguchi}}, \bibinfo {author} {\bibfnamefont {K.}~\bibnamefont {Kiuchi}},
  \bibinfo {author} {\bibfnamefont {K.}~\bibnamefont {Kyutoku}},\ and\ \bibinfo
  {author} {\bibfnamefont {M.}~\bibnamefont {Shibata}},\ }\bibfield  {title}
  {\bibinfo {title} {{Gravitational waves and neutrino emission from the merger
  of binary neutron stars}},\ }\href
  {https://doi.org/10.1103/PhysRevLett.107.051102} {\bibfield  {journal}
  {\bibinfo  {journal} {Phys. Rev. Lett.}\ }\textbf {\bibinfo {volume} {107}},\
  \bibinfo {pages} {051102} (\bibinfo {year} {2011})},\ \Eprint
  {https://arxiv.org/abs/1105.2125} {arXiv:1105.2125 [gr-qc]} \BibitemShut
  {NoStop}%
\bibitem [{\citenamefont {Foucart}(2018)}]{Foucart:2017mbt}%
  \BibitemOpen
  \bibfield  {author} {\bibinfo {author} {\bibfnamefont {F.}~\bibnamefont
  {Foucart}},\ }\bibfield  {title} {\bibinfo {title} {{Monte Carlo closure for
  moment-based transport schemes in general relativistic radiation hydrodynamic
  simulations}},\ }\href {https://doi.org/10.1093/mnras/sty108} {\bibfield
  {journal} {\bibinfo  {journal} {Mon. Not. Roy. Astron. Soc.}\ }\textbf
  {\bibinfo {volume} {475}},\ \bibinfo {pages} {4186} (\bibinfo {year}
  {2018})},\ \Eprint {https://arxiv.org/abs/1708.08452} {arXiv:1708.08452
  [astro-ph.HE]} \BibitemShut {NoStop}%
\bibitem [{\citenamefont {Foucart}\ \emph {et~al.}(2018)\citenamefont
  {Foucart}, \citenamefont {Duez}, \citenamefont {Kidder}, \citenamefont
  {Nguyen}, \citenamefont {Pfeiffer},\ and\ \citenamefont
  {Scheel}}]{Foucart:2018gis}%
  \BibitemOpen
  \bibfield  {author} {\bibinfo {author} {\bibfnamefont {F.}~\bibnamefont
  {Foucart}}, \bibinfo {author} {\bibfnamefont {M.~D.}\ \bibnamefont {Duez}},
  \bibinfo {author} {\bibfnamefont {L.~E.}\ \bibnamefont {Kidder}}, \bibinfo
  {author} {\bibfnamefont {R.}~\bibnamefont {Nguyen}}, \bibinfo {author}
  {\bibfnamefont {H.~P.}\ \bibnamefont {Pfeiffer}},\ and\ \bibinfo {author}
  {\bibfnamefont {M.~A.}\ \bibnamefont {Scheel}},\ }\bibfield  {title}
  {\bibinfo {title} {{Evaluating radiation transport errors in merger
  simulations using a Monte Carlo algorithm}},\ }\href
  {https://doi.org/10.1103/PhysRevD.98.063007} {\bibfield  {journal} {\bibinfo
  {journal} {Phys. Rev. D}\ }\textbf {\bibinfo {volume} {98}},\ \bibinfo
  {pages} {063007} (\bibinfo {year} {2018})},\ \Eprint
  {https://arxiv.org/abs/1806.02349} {arXiv:1806.02349 [astro-ph.HE]}
  \BibitemShut {NoStop}%
\bibitem [{\citenamefont {Foucart}\ \emph {et~al.}(2020)\citenamefont
  {Foucart}, \citenamefont {Duez}, \citenamefont {Hebert}, \citenamefont
  {Kidder}, \citenamefont {Pfeiffer},\ and\ \citenamefont
  {Scheel}}]{Foucart:2020qjb}%
  \BibitemOpen
  \bibfield  {author} {\bibinfo {author} {\bibfnamefont {F.}~\bibnamefont
  {Foucart}}, \bibinfo {author} {\bibfnamefont {M.~D.}\ \bibnamefont {Duez}},
  \bibinfo {author} {\bibfnamefont {F.}~\bibnamefont {Hebert}}, \bibinfo
  {author} {\bibfnamefont {L.~E.}\ \bibnamefont {Kidder}}, \bibinfo {author}
  {\bibfnamefont {H.~P.}\ \bibnamefont {Pfeiffer}},\ and\ \bibinfo {author}
  {\bibfnamefont {M.~A.}\ \bibnamefont {Scheel}},\ }\bibfield  {title}
  {\bibinfo {title} {{Monte-Carlo neutrino transport in neutron star merger
  simulations}},\ }\href {https://doi.org/10.3847/2041-8213/abbb87} {\bibfield
  {journal} {\bibinfo  {journal} {Astrophys. J. Lett.}\ }\textbf {\bibinfo
  {volume} {902}},\ \bibinfo {pages} {L27} (\bibinfo {year} {2020})},\ \Eprint
  {https://arxiv.org/abs/2008.08089} {arXiv:2008.08089 [astro-ph.HE]}
  \BibitemShut {NoStop}%
\bibitem [{\citenamefont {Most}\ \emph {et~al.}(2019)\citenamefont {Most},
  \citenamefont {Papenfort}, \citenamefont {Dexheimer}, \citenamefont
  {Hanauske}, \citenamefont {Schramm}, \citenamefont {St\"ocker},\ and\
  \citenamefont {Rezzolla}}]{Most:2018eaw}%
  \BibitemOpen
  \bibfield  {author} {\bibinfo {author} {\bibfnamefont {E.~R.}\ \bibnamefont
  {Most}}, \bibinfo {author} {\bibfnamefont {L.~J.}\ \bibnamefont {Papenfort}},
  \bibinfo {author} {\bibfnamefont {V.}~\bibnamefont {Dexheimer}}, \bibinfo
  {author} {\bibfnamefont {M.}~\bibnamefont {Hanauske}}, \bibinfo {author}
  {\bibfnamefont {S.}~\bibnamefont {Schramm}}, \bibinfo {author} {\bibfnamefont
  {H.}~\bibnamefont {St\"ocker}},\ and\ \bibinfo {author} {\bibfnamefont
  {L.}~\bibnamefont {Rezzolla}},\ }\bibfield  {title} {\bibinfo {title}
  {{Signatures of quark-hadron phase transitions in general-relativistic
  neutron-star mergers}},\ }\href
  {https://doi.org/10.1103/PhysRevLett.122.061101} {\bibfield  {journal}
  {\bibinfo  {journal} {Phys. Rev. Lett.}\ }\textbf {\bibinfo {volume} {122}},\
  \bibinfo {pages} {061101} (\bibinfo {year} {2019})},\ \Eprint
  {https://arxiv.org/abs/1807.03684} {arXiv:1807.03684 [astro-ph.HE]}
  \BibitemShut {NoStop}%
\bibitem [{\citenamefont {Bauswein}\ \emph {et~al.}(2019)\citenamefont
  {Bauswein}, \citenamefont {Bastian}, \citenamefont {Blaschke}, \citenamefont
  {Chatziioannou}, \citenamefont {Clark}, \citenamefont {Fischer},\ and\
  \citenamefont {Oertel}}]{Bauswein:2018bma}%
  \BibitemOpen
  \bibfield  {author} {\bibinfo {author} {\bibfnamefont {A.}~\bibnamefont
  {Bauswein}}, \bibinfo {author} {\bibfnamefont {N.-U.~F.}\ \bibnamefont
  {Bastian}}, \bibinfo {author} {\bibfnamefont {D.~B.}\ \bibnamefont
  {Blaschke}}, \bibinfo {author} {\bibfnamefont {K.}~\bibnamefont
  {Chatziioannou}}, \bibinfo {author} {\bibfnamefont {J.~A.}\ \bibnamefont
  {Clark}}, \bibinfo {author} {\bibfnamefont {T.}~\bibnamefont {Fischer}},\
  and\ \bibinfo {author} {\bibfnamefont {M.}~\bibnamefont {Oertel}},\
  }\bibfield  {title} {\bibinfo {title} {{Identifying a first-order phase
  transition in neutron star mergers through gravitational waves}},\ }\href
  {https://doi.org/10.1103/PhysRevLett.122.061102} {\bibfield  {journal}
  {\bibinfo  {journal} {Phys. Rev. Lett.}\ }\textbf {\bibinfo {volume} {122}},\
  \bibinfo {pages} {061102} (\bibinfo {year} {2019})},\ \Eprint
  {https://arxiv.org/abs/1809.01116} {arXiv:1809.01116 [astro-ph.HE]}
  \BibitemShut {NoStop}%
\bibitem [{\citenamefont {Weih}\ \emph {et~al.}(2020)\citenamefont {Weih},
  \citenamefont {Hanauske},\ and\ \citenamefont {Rezzolla}}]{Weih:2019xvw}%
  \BibitemOpen
  \bibfield  {author} {\bibinfo {author} {\bibfnamefont {L.~R.}\ \bibnamefont
  {Weih}}, \bibinfo {author} {\bibfnamefont {M.}~\bibnamefont {Hanauske}},\
  and\ \bibinfo {author} {\bibfnamefont {L.}~\bibnamefont {Rezzolla}},\
  }\bibfield  {title} {\bibinfo {title} {{Postmerger Gravitational-Wave
  Signatures of Phase Transitions in Binary Mergers}},\ }\href
  {https://doi.org/10.1103/PhysRevLett.124.171103} {\bibfield  {journal}
  {\bibinfo  {journal} {Phys. Rev. Lett.}\ }\textbf {\bibinfo {volume} {124}},\
  \bibinfo {pages} {171103} (\bibinfo {year} {2020})},\ \Eprint
  {https://arxiv.org/abs/1912.09340} {arXiv:1912.09340 [gr-qc]} \BibitemShut
  {NoStop}%
\bibitem [{\citenamefont {Blacker}\ \emph {et~al.}(2020)\citenamefont
  {Blacker}, \citenamefont {Bastian}, \citenamefont {Bauswein}, \citenamefont
  {Blaschke}, \citenamefont {Fischer}, \citenamefont {Oertel}, \citenamefont
  {Soultanis},\ and\ \citenamefont {Typel}}]{Blacker:2020nlq}%
  \BibitemOpen
  \bibfield  {author} {\bibinfo {author} {\bibfnamefont {S.}~\bibnamefont
  {Blacker}}, \bibinfo {author} {\bibfnamefont {N.-U.~F.}\ \bibnamefont
  {Bastian}}, \bibinfo {author} {\bibfnamefont {A.}~\bibnamefont {Bauswein}},
  \bibinfo {author} {\bibfnamefont {D.~B.}\ \bibnamefont {Blaschke}}, \bibinfo
  {author} {\bibfnamefont {T.}~\bibnamefont {Fischer}}, \bibinfo {author}
  {\bibfnamefont {M.}~\bibnamefont {Oertel}}, \bibinfo {author} {\bibfnamefont
  {T.}~\bibnamefont {Soultanis}},\ and\ \bibinfo {author} {\bibfnamefont
  {S.}~\bibnamefont {Typel}},\ }\bibfield  {title} {\bibinfo {title}
  {{Constraining the onset density of the hadron-quark phase transition with
  gravitational-wave observations}},\ }\href
  {https://doi.org/10.1103/PhysRevD.102.123023} {\bibfield  {journal} {\bibinfo
   {journal} {Phys. Rev. D}\ }\textbf {\bibinfo {volume} {102}},\ \bibinfo
  {pages} {123023} (\bibinfo {year} {2020})},\ \Eprint
  {https://arxiv.org/abs/2006.03789} {arXiv:2006.03789 [astro-ph.HE]}
  \BibitemShut {NoStop}%
\bibitem [{\citenamefont {Bauswein}\ and\ \citenamefont
  {Blacker}(2020)}]{Bauswein:2020ggy}%
  \BibitemOpen
  \bibfield  {author} {\bibinfo {author} {\bibfnamefont {A.}~\bibnamefont
  {Bauswein}}\ and\ \bibinfo {author} {\bibfnamefont {S.}~\bibnamefont
  {Blacker}},\ }\bibfield  {title} {\bibinfo {title} {{Impact of quark
  deconfinement in neutron star mergers and hybrid star mergers}},\ }\href
  {https://doi.org/10.1140/epjst/e2020-000138-7} {\bibfield  {journal}
  {\bibinfo  {journal} {Eur. Phys. J. ST}\ }\textbf {\bibinfo {volume} {229}},\
  \bibinfo {pages} {3595} (\bibinfo {year} {2020})},\ \Eprint
  {https://arxiv.org/abs/2006.16183} {arXiv:2006.16183 [astro-ph.HE]}
  \BibitemShut {NoStop}%
\bibitem [{\citenamefont {Huang}\ \emph {et~al.}(2022)\citenamefont {Huang},
  \citenamefont {Baiotti}, \citenamefont {Kojo}, \citenamefont {Takami},
  \citenamefont {Sotani}, \citenamefont {Togashi}, \citenamefont {Hatsuda},
  \citenamefont {Nagataki},\ and\ \citenamefont {Fan}}]{Huang:2022mqp}%
  \BibitemOpen
  \bibfield  {author} {\bibinfo {author} {\bibfnamefont {Y.-J.}\ \bibnamefont
  {Huang}}, \bibinfo {author} {\bibfnamefont {L.}~\bibnamefont {Baiotti}},
  \bibinfo {author} {\bibfnamefont {T.}~\bibnamefont {Kojo}}, \bibinfo {author}
  {\bibfnamefont {K.}~\bibnamefont {Takami}}, \bibinfo {author} {\bibfnamefont
  {H.}~\bibnamefont {Sotani}}, \bibinfo {author} {\bibfnamefont
  {H.}~\bibnamefont {Togashi}}, \bibinfo {author} {\bibfnamefont
  {T.}~\bibnamefont {Hatsuda}}, \bibinfo {author} {\bibfnamefont
  {S.}~\bibnamefont {Nagataki}},\ and\ \bibinfo {author} {\bibfnamefont
  {Y.-Z.}\ \bibnamefont {Fan}},\ }\bibfield  {title} {\bibinfo {title} {{Merger
  and post-merger of binary neutron stars with a quark-hadron crossover
  equation of state}},\ }\href@noop {} {\  (\bibinfo {year} {2022})},\ \Eprint
  {https://arxiv.org/abs/2203.04528} {arXiv:2203.04528 [astro-ph.HE]}
  \BibitemShut {NoStop}%
\bibitem [{\citenamefont {Glendenning}(1992)}]{Glendenning:1992vb}%
  \BibitemOpen
  \bibfield  {author} {\bibinfo {author} {\bibfnamefont {N.~K.}\ \bibnamefont
  {Glendenning}},\ }\bibfield  {title} {\bibinfo {title} {{First order phase
  transitions with more than one conserved charge: Consequences for neutron
  stars}},\ }\href {https://doi.org/10.1103/PhysRevD.46.1274} {\bibfield
  {journal} {\bibinfo  {journal} {Phys. Rev. D}\ }\textbf {\bibinfo {volume}
  {46}},\ \bibinfo {pages} {1274} (\bibinfo {year} {1992})}\BibitemShut
  {NoStop}%
\bibitem [{\citenamefont {Alford}(2001)}]{Alford:2001dt}%
  \BibitemOpen
  \bibfield  {author} {\bibinfo {author} {\bibfnamefont {M.~G.}\ \bibnamefont
  {Alford}},\ }\bibfield  {title} {\bibinfo {title} {{Color superconducting
  quark matter}},\ }\href
  {https://doi.org/10.1146/annurev.nucl.51.101701.132449} {\bibfield  {journal}
  {\bibinfo  {journal} {Ann. Rev. Nucl. Part. Sci.}\ }\textbf {\bibinfo
  {volume} {51}},\ \bibinfo {pages} {131} (\bibinfo {year} {2001})},\ \Eprint
  {https://arxiv.org/abs/hep-ph/0102047} {arXiv:hep-ph/0102047} \BibitemShut
  {NoStop}%
\bibitem [{\citenamefont {Alford}\ \emph {et~al.}(2005)\citenamefont {Alford},
  \citenamefont {Braby}, \citenamefont {Paris},\ and\ \citenamefont
  {Reddy}}]{Alford:2004pf}%
  \BibitemOpen
  \bibfield  {author} {\bibinfo {author} {\bibfnamefont {M.}~\bibnamefont
  {Alford}}, \bibinfo {author} {\bibfnamefont {M.}~\bibnamefont {Braby}},
  \bibinfo {author} {\bibfnamefont {M.~W.}\ \bibnamefont {Paris}},\ and\
  \bibinfo {author} {\bibfnamefont {S.}~\bibnamefont {Reddy}},\ }\bibfield
  {title} {\bibinfo {title} {{Hybrid stars that masquerade as neutron stars}},\
  }\href {https://doi.org/10.1086/430902} {\bibfield  {journal} {\bibinfo
  {journal} {Astrophys. J.}\ }\textbf {\bibinfo {volume} {629}},\ \bibinfo
  {pages} {969} (\bibinfo {year} {2005})},\ \Eprint
  {https://arxiv.org/abs/nucl-th/0411016} {arXiv:nucl-th/0411016} \BibitemShut
  {NoStop}%
\bibitem [{\citenamefont {Benic}\ \emph {et~al.}(2015)\citenamefont {Benic},
  \citenamefont {Blaschke}, \citenamefont {Alvarez-Castillo}, \citenamefont
  {Fischer},\ and\ \citenamefont {Typel}}]{Benic:2014jia}%
  \BibitemOpen
  \bibfield  {author} {\bibinfo {author} {\bibfnamefont {S.}~\bibnamefont
  {Benic}}, \bibinfo {author} {\bibfnamefont {D.}~\bibnamefont {Blaschke}},
  \bibinfo {author} {\bibfnamefont {D.~E.}\ \bibnamefont {Alvarez-Castillo}},
  \bibinfo {author} {\bibfnamefont {T.}~\bibnamefont {Fischer}},\ and\ \bibinfo
  {author} {\bibfnamefont {S.}~\bibnamefont {Typel}},\ }\bibfield  {title}
  {\bibinfo {title} {{A new quark-hadron hybrid equation of state for
  astrophysics - I. High-mass twin compact stars}},\ }\href
  {https://doi.org/10.1051/0004-6361/201425318} {\bibfield  {journal} {\bibinfo
   {journal} {Astron. Astrophys.}\ }\textbf {\bibinfo {volume} {577}},\
  \bibinfo {pages} {A40} (\bibinfo {year} {2015})},\ \Eprint
  {https://arxiv.org/abs/1411.2856} {arXiv:1411.2856 [astro-ph.HE]}
  \BibitemShut {NoStop}%
\bibitem [{\citenamefont {Montana}\ \emph {et~al.}(2019)\citenamefont
  {Montana}, \citenamefont {Tolos}, \citenamefont {Hanauske},\ and\
  \citenamefont {Rezzolla}}]{Montana:2018bkb}%
  \BibitemOpen
  \bibfield  {author} {\bibinfo {author} {\bibfnamefont {G.}~\bibnamefont
  {Montana}}, \bibinfo {author} {\bibfnamefont {L.}~\bibnamefont {Tolos}},
  \bibinfo {author} {\bibfnamefont {M.}~\bibnamefont {Hanauske}},\ and\
  \bibinfo {author} {\bibfnamefont {L.}~\bibnamefont {Rezzolla}},\ }\bibfield
  {title} {\bibinfo {title} {{Constraining twin stars with GW170817}},\ }\href
  {https://doi.org/10.1103/PhysRevD.99.103009} {\bibfield  {journal} {\bibinfo
  {journal} {Phys. Rev. D}\ }\textbf {\bibinfo {volume} {99}},\ \bibinfo
  {pages} {103009} (\bibinfo {year} {2019})},\ \Eprint
  {https://arxiv.org/abs/1811.10929} {arXiv:1811.10929 [astro-ph.HE]}
  \BibitemShut {NoStop}%
\bibitem [{\citenamefont {Gieg}\ \emph {et~al.}(2019)\citenamefont {Gieg},
  \citenamefont {Dietrich},\ and\ \citenamefont {Ujevic}}]{Gieg:2019yzq}%
  \BibitemOpen
  \bibfield  {author} {\bibinfo {author} {\bibfnamefont {H.}~\bibnamefont
  {Gieg}}, \bibinfo {author} {\bibfnamefont {T.}~\bibnamefont {Dietrich}},\
  and\ \bibinfo {author} {\bibfnamefont {M.}~\bibnamefont {Ujevic}},\
  }\bibfield  {title} {\bibinfo {title} {{Simulating Binary Neutron Stars with
  Hybrid Equation of States: Gravitational Waves, Electromagnetic Signatures,
  and Challenges for Numerical Relativity}},\ }\href
  {https://doi.org/10.3390/particles2030023} {\bibfield  {journal} {\bibinfo
  {journal} {Particles}\ }\textbf {\bibinfo {volume} {2}},\ \bibinfo {pages}
  {365} (\bibinfo {year} {2019})},\ \Eprint {https://arxiv.org/abs/1908.03135}
  {arXiv:1908.03135 [gr-qc]} \BibitemShut {NoStop}%
\bibitem [{\citenamefont {Pang}\ \emph {et~al.}(2020)\citenamefont {Pang},
  \citenamefont {Dietrich}, \citenamefont {Tews},\ and\ \citenamefont {Van
  Den~Broeck}}]{Pang:2020ilf}%
  \BibitemOpen
  \bibfield  {author} {\bibinfo {author} {\bibfnamefont {P.~T.~H.}\
  \bibnamefont {Pang}}, \bibinfo {author} {\bibfnamefont {T.}~\bibnamefont
  {Dietrich}}, \bibinfo {author} {\bibfnamefont {I.}~\bibnamefont {Tews}},\
  and\ \bibinfo {author} {\bibfnamefont {C.}~\bibnamefont {Van Den~Broeck}},\
  }\bibfield  {title} {\bibinfo {title} {{Parameter estimation for strong phase
  transitions in supranuclear matter using gravitational-wave astronomy}},\
  }\href {https://doi.org/10.1103/PhysRevResearch.2.033514} {\bibfield
  {journal} {\bibinfo  {journal} {Phys. Rev. Res.}\ }\textbf {\bibinfo {volume}
  {2}},\ \bibinfo {pages} {033514} (\bibinfo {year} {2020})},\ \Eprint
  {https://arxiv.org/abs/2006.14936} {arXiv:2006.14936 [astro-ph.HE]}
  \BibitemShut {NoStop}%
\bibitem [{\citenamefont {Li}\ \emph {et~al.}(2018)\citenamefont {Li},
  \citenamefont {Yan}, \citenamefont {Geng}, \citenamefont {Huang},\ and\
  \citenamefont {Zong}}]{Li:2018ayl}%
  \BibitemOpen
  \bibfield  {author} {\bibinfo {author} {\bibfnamefont {C.-M.}\ \bibnamefont
  {Li}}, \bibinfo {author} {\bibfnamefont {Y.}~\bibnamefont {Yan}}, \bibinfo
  {author} {\bibfnamefont {J.-J.}\ \bibnamefont {Geng}}, \bibinfo {author}
  {\bibfnamefont {Y.-F.}\ \bibnamefont {Huang}},\ and\ \bibinfo {author}
  {\bibfnamefont {H.-S.}\ \bibnamefont {Zong}},\ }\bibfield  {title} {\bibinfo
  {title} {{Constraints on the hybrid equation of state with a crossover
  hadron-quark phase transition in the light of GW170817}},\ }\href
  {https://doi.org/10.1103/PhysRevD.98.083013} {\bibfield  {journal} {\bibinfo
  {journal} {Phys. Rev. D}\ }\textbf {\bibinfo {volume} {98}},\ \bibinfo
  {pages} {083013} (\bibinfo {year} {2018})},\ \Eprint
  {https://arxiv.org/abs/1808.02601} {arXiv:1808.02601 [nucl-th]} \BibitemShut
  {NoStop}%
\bibitem [{\citenamefont {Christian}\ \emph {et~al.}(2019)\citenamefont
  {Christian}, \citenamefont {Zacchi},\ and\ \citenamefont
  {Schaffner-Bielich}}]{Christian:2018jyd}%
  \BibitemOpen
  \bibfield  {author} {\bibinfo {author} {\bibfnamefont {J.-E.}\ \bibnamefont
  {Christian}}, \bibinfo {author} {\bibfnamefont {A.}~\bibnamefont {Zacchi}},\
  and\ \bibinfo {author} {\bibfnamefont {J.}~\bibnamefont
  {Schaffner-Bielich}},\ }\bibfield  {title} {\bibinfo {title} {{Signals in the
  tidal deformability for phase transitions in compact stars with constraints
  from GW170817}},\ }\href {https://doi.org/10.1103/PhysRevD.99.023009}
  {\bibfield  {journal} {\bibinfo  {journal} {Phys. Rev. D}\ }\textbf {\bibinfo
  {volume} {99}},\ \bibinfo {pages} {023009} (\bibinfo {year} {2019})},\
  \Eprint {https://arxiv.org/abs/1809.03333} {arXiv:1809.03333 [astro-ph.HE]}
  \BibitemShut {NoStop}%
\bibitem [{\citenamefont {Kashyap}\ \emph {et~al.}(2022)\citenamefont {Kashyap}
  \emph {et~al.}}]{Kashyap:2021wzs}%
  \BibitemOpen
  \bibfield  {author} {\bibinfo {author} {\bibfnamefont {R.}~\bibnamefont
  {Kashyap}} \emph {et~al.},\ }\bibfield  {title} {\bibinfo {title} {{Numerical
  relativity simulations of prompt collapse mergers: Threshold mass and
  phenomenological constraints on neutron star properties after GW170817}},\
  }\href {https://doi.org/10.1103/PhysRevD.105.103022} {\bibfield  {journal}
  {\bibinfo  {journal} {Phys. Rev. D}\ }\textbf {\bibinfo {volume} {105}},\
  \bibinfo {pages} {103022} (\bibinfo {year} {2022})},\ \Eprint
  {https://arxiv.org/abs/2111.05183} {arXiv:2111.05183 [astro-ph.HE]}
  \BibitemShut {NoStop}%
\bibitem [{\citenamefont {Chatziioannou}\ and\ \citenamefont
  {Han}(2020)}]{Chatziioannou:2019yko}%
  \BibitemOpen
  \bibfield  {author} {\bibinfo {author} {\bibfnamefont {K.}~\bibnamefont
  {Chatziioannou}}\ and\ \bibinfo {author} {\bibfnamefont {S.}~\bibnamefont
  {Han}},\ }\bibfield  {title} {\bibinfo {title} {{Studying strong phase
  transitions in neutron stars with gravitational waves}},\ }\href
  {https://doi.org/10.1103/PhysRevD.101.044019} {\bibfield  {journal} {\bibinfo
   {journal} {Phys. Rev. D}\ }\textbf {\bibinfo {volume} {101}},\ \bibinfo
  {pages} {044019} (\bibinfo {year} {2020})},\ \Eprint
  {https://arxiv.org/abs/1911.07091} {arXiv:1911.07091 [gr-qc]} \BibitemShut
  {NoStop}%
\bibitem [{\citenamefont {Haque}\ \emph {et~al.}(2022)\citenamefont {Haque},
  \citenamefont {Mallick},\ and\ \citenamefont {Thakur}}]{Haque:2022dsc}%
  \BibitemOpen
  \bibfield  {author} {\bibinfo {author} {\bibfnamefont {S.}~\bibnamefont
  {Haque}}, \bibinfo {author} {\bibfnamefont {R.}~\bibnamefont {Mallick}},\
  and\ \bibinfo {author} {\bibfnamefont {S.~K.}\ \bibnamefont {Thakur}},\
  }\bibfield  {title} {\bibinfo {title} {{Binary neutron star mergers and the
  effect of onset of phase transition on gravitational wave signals}},\
  }\href@noop {} {\  (\bibinfo {year} {2022})},\ \Eprint
  {https://arxiv.org/abs/2207.14485} {arXiv:2207.14485 [astro-ph.HE]}
  \BibitemShut {NoStop}%
\bibitem [{\citenamefont {Tootle}\ \emph {et~al.}(2022)\citenamefont {Tootle},
  \citenamefont {Ecker}, \citenamefont {Topolski}, \citenamefont {Demircik},
  \citenamefont {J\"arvinen},\ and\ \citenamefont {Rezzolla}}]{Tootle:2022pvd}%
  \BibitemOpen
  \bibfield  {author} {\bibinfo {author} {\bibfnamefont {S.}~\bibnamefont
  {Tootle}}, \bibinfo {author} {\bibfnamefont {C.}~\bibnamefont {Ecker}},
  \bibinfo {author} {\bibfnamefont {K.}~\bibnamefont {Topolski}}, \bibinfo
  {author} {\bibfnamefont {T.}~\bibnamefont {Demircik}}, \bibinfo {author}
  {\bibfnamefont {M.}~\bibnamefont {J\"arvinen}},\ and\ \bibinfo {author}
  {\bibfnamefont {L.}~\bibnamefont {Rezzolla}},\ }\bibfield  {title} {\bibinfo
  {title} {{Quark formation and phenomenology in binary neutron-star mergers
  using V-QCD}},\ }\href@noop {} {\  (\bibinfo {year} {2022})},\ \Eprint
  {https://arxiv.org/abs/2205.05691} {arXiv:2205.05691 [astro-ph.HE]}
  \BibitemShut {NoStop}%
\bibitem [{\citenamefont {Fujimoto}\ \emph {et~al.}(2022)\citenamefont
  {Fujimoto}, \citenamefont {Fukushima}, \citenamefont {Hotokezaka},\ and\
  \citenamefont {Kyutoku}}]{Fujimoto:2022xhv}%
  \BibitemOpen
  \bibfield  {author} {\bibinfo {author} {\bibfnamefont {Y.}~\bibnamefont
  {Fujimoto}}, \bibinfo {author} {\bibfnamefont {K.}~\bibnamefont {Fukushima}},
  \bibinfo {author} {\bibfnamefont {K.}~\bibnamefont {Hotokezaka}},\ and\
  \bibinfo {author} {\bibfnamefont {K.}~\bibnamefont {Kyutoku}},\ }\bibfield
  {title} {\bibinfo {title} {{Gravitational Wave Signal for Quark Matter with
  Realistic Phase Transition}},\ }\href@noop {} {\  (\bibinfo {year} {2022})},\
  \Eprint {https://arxiv.org/abs/2205.03882} {arXiv:2205.03882 [astro-ph.HE]}
  \BibitemShut {NoStop}%
\bibitem [{\citenamefont {Prakash}\ \emph {et~al.}(2021)\citenamefont
  {Prakash}, \citenamefont {Radice}, \citenamefont {Logoteta}, \citenamefont
  {Perego}, \citenamefont {Nedora}, \citenamefont {Bombaci}, \citenamefont
  {Kashyap}, \citenamefont {Bernuzzi},\ and\ \citenamefont
  {Endrizzi}}]{Prakash:2021wpz}%
  \BibitemOpen
  \bibfield  {author} {\bibinfo {author} {\bibfnamefont {A.}~\bibnamefont
  {Prakash}}, \bibinfo {author} {\bibfnamefont {D.}~\bibnamefont {Radice}},
  \bibinfo {author} {\bibfnamefont {D.}~\bibnamefont {Logoteta}}, \bibinfo
  {author} {\bibfnamefont {A.}~\bibnamefont {Perego}}, \bibinfo {author}
  {\bibfnamefont {V.}~\bibnamefont {Nedora}}, \bibinfo {author} {\bibfnamefont
  {I.}~\bibnamefont {Bombaci}}, \bibinfo {author} {\bibfnamefont
  {R.}~\bibnamefont {Kashyap}}, \bibinfo {author} {\bibfnamefont
  {S.}~\bibnamefont {Bernuzzi}},\ and\ \bibinfo {author} {\bibfnamefont
  {A.}~\bibnamefont {Endrizzi}},\ }\bibfield  {title} {\bibinfo {title}
  {{Signatures of deconfined quark phases in binary neutron star mergers}},\
  }\href {https://doi.org/10.1103/PhysRevD.104.083029} {\bibfield  {journal}
  {\bibinfo  {journal} {Phys. Rev. D}\ }\textbf {\bibinfo {volume} {104}},\
  \bibinfo {pages} {083029} (\bibinfo {year} {2021})},\ \Eprint
  {https://arxiv.org/abs/2106.07885} {arXiv:2106.07885 [astro-ph.HE]}
  \BibitemShut {NoStop}%
\bibitem [{\citenamefont {Liebling}\ \emph {et~al.}(2021)\citenamefont
  {Liebling}, \citenamefont {Palenzuela},\ and\ \citenamefont
  {Lehner}}]{Liebling:2020dhf}%
  \BibitemOpen
  \bibfield  {author} {\bibinfo {author} {\bibfnamefont {S.~L.}\ \bibnamefont
  {Liebling}}, \bibinfo {author} {\bibfnamefont {C.}~\bibnamefont
  {Palenzuela}},\ and\ \bibinfo {author} {\bibfnamefont {L.}~\bibnamefont
  {Lehner}},\ }\bibfield  {title} {\bibinfo {title} {{Effects of High Density
  Phase Transitions on Neutron Star Dynamics}},\ }\href
  {https://doi.org/10.1088/1361-6382/abf898} {\bibfield  {journal} {\bibinfo
  {journal} {Class. Quant. Grav.}\ }\textbf {\bibinfo {volume} {38}},\ \bibinfo
  {pages} {115007} (\bibinfo {year} {2021})},\ \Eprint
  {https://arxiv.org/abs/2010.12567} {arXiv:2010.12567 [gr-qc]} \BibitemShut
  {NoStop}%
\bibitem [{\citenamefont {Moldenhauer}\ \emph {et~al.}(2014)\citenamefont
  {Moldenhauer}, \citenamefont {Markakis}, \citenamefont {Johnson-McDaniel},
  \citenamefont {Tichy},\ and\ \citenamefont
  {Br\"ugmann}}]{Moldenhauer:2014yaa}%
  \BibitemOpen
  \bibfield  {author} {\bibinfo {author} {\bibfnamefont {N.}~\bibnamefont
  {Moldenhauer}}, \bibinfo {author} {\bibfnamefont {C.~M.}\ \bibnamefont
  {Markakis}}, \bibinfo {author} {\bibfnamefont {N.~K.}\ \bibnamefont
  {Johnson-McDaniel}}, \bibinfo {author} {\bibfnamefont {W.}~\bibnamefont
  {Tichy}},\ and\ \bibinfo {author} {\bibfnamefont {B.}~\bibnamefont
  {Br\"ugmann}},\ }\bibfield  {title} {\bibinfo {title} {{Initial data for
  binary neutron stars with adjustable eccentricity}},\ }\href
  {https://doi.org/10.1103/PhysRevD.90.084043} {\bibfield  {journal} {\bibinfo
  {journal} {Phys. Rev. D}\ }\textbf {\bibinfo {volume} {90}},\ \bibinfo
  {pages} {084043} (\bibinfo {year} {2014})},\ \Eprint
  {https://arxiv.org/abs/1408.4136} {arXiv:1408.4136 [gr-qc]} \BibitemShut
  {NoStop}%
\bibitem [{\citenamefont {Dietrich}\ \emph
  {et~al.}(2015{\natexlab{a}})\citenamefont {Dietrich}, \citenamefont
  {Moldenhauer}, \citenamefont {Johnson-McDaniel}, \citenamefont {Bernuzzi},
  \citenamefont {Markakis}, \citenamefont {Br\"ugmann},\ and\ \citenamefont
  {Tichy}}]{Dietrich:2015pxa}%
  \BibitemOpen
  \bibfield  {author} {\bibinfo {author} {\bibfnamefont {T.}~\bibnamefont
  {Dietrich}}, \bibinfo {author} {\bibfnamefont {N.}~\bibnamefont
  {Moldenhauer}}, \bibinfo {author} {\bibfnamefont {N.~K.}\ \bibnamefont
  {Johnson-McDaniel}}, \bibinfo {author} {\bibfnamefont {S.}~\bibnamefont
  {Bernuzzi}}, \bibinfo {author} {\bibfnamefont {C.~M.}\ \bibnamefont
  {Markakis}}, \bibinfo {author} {\bibfnamefont {B.}~\bibnamefont
  {Br\"ugmann}},\ and\ \bibinfo {author} {\bibfnamefont {W.}~\bibnamefont
  {Tichy}},\ }\bibfield  {title} {\bibinfo {title} {{Binary Neutron Stars with
  Generic Spin, Eccentricity, Mass ratio, and Compactness - Quasi-equilibrium
  Sequences and First Evolutions}},\ }\href
  {https://doi.org/10.1103/PhysRevD.92.124007} {\bibfield  {journal} {\bibinfo
  {journal} {Phys. Rev. D}\ }\textbf {\bibinfo {volume} {92}},\ \bibinfo
  {pages} {124007} (\bibinfo {year} {2015}{\natexlab{a}})},\ \Eprint
  {https://arxiv.org/abs/1507.07100} {arXiv:1507.07100 [gr-qc]} \BibitemShut
  {NoStop}%
\bibitem [{\citenamefont {Lynn}\ \emph {et~al.}(2016)\citenamefont {Lynn},
  \citenamefont {Tews}, \citenamefont {Carlson}, \citenamefont {Gandolfi},
  \citenamefont {Gezerlis}, \citenamefont {Schmidt},\ and\ \citenamefont
  {Schwenk}}]{Lynn:2015jua}%
  \BibitemOpen
  \bibfield  {author} {\bibinfo {author} {\bibfnamefont {J.~E.}\ \bibnamefont
  {Lynn}}, \bibinfo {author} {\bibfnamefont {I.}~\bibnamefont {Tews}}, \bibinfo
  {author} {\bibfnamefont {J.}~\bibnamefont {Carlson}}, \bibinfo {author}
  {\bibfnamefont {S.}~\bibnamefont {Gandolfi}}, \bibinfo {author}
  {\bibfnamefont {A.}~\bibnamefont {Gezerlis}}, \bibinfo {author}
  {\bibfnamefont {K.~E.}\ \bibnamefont {Schmidt}},\ and\ \bibinfo {author}
  {\bibfnamefont {A.}~\bibnamefont {Schwenk}},\ }\bibfield  {title} {\bibinfo
  {title} {{Chiral Three-Nucleon Interactions in Light Nuclei, Neutron-$\alpha$
  Scattering, and Neutron Matter}},\ }\href
  {https://doi.org/10.1103/PhysRevLett.116.062501} {\bibfield  {journal}
  {\bibinfo  {journal} {Phys. Rev. Lett.}\ }\textbf {\bibinfo {volume} {116}},\
  \bibinfo {pages} {062501} (\bibinfo {year} {2016})},\ \Eprint
  {https://arxiv.org/abs/1509.03470} {arXiv:1509.03470 [nucl-th]} \BibitemShut
  {NoStop}%
\bibitem [{\citenamefont {Tews}\ \emph {et~al.}(2018)\citenamefont {Tews},
  \citenamefont {Carlson}, \citenamefont {Gandolfi},\ and\ \citenamefont
  {Reddy}}]{Tews:2018kmu}%
  \BibitemOpen
  \bibfield  {author} {\bibinfo {author} {\bibfnamefont {I.}~\bibnamefont
  {Tews}}, \bibinfo {author} {\bibfnamefont {J.}~\bibnamefont {Carlson}},
  \bibinfo {author} {\bibfnamefont {S.}~\bibnamefont {Gandolfi}},\ and\
  \bibinfo {author} {\bibfnamefont {S.}~\bibnamefont {Reddy}},\ }\bibfield
  {title} {\bibinfo {title} {{Constraining the speed of sound inside neutron
  stars with chiral effective field theory interactions and observations}},\
  }\href {https://doi.org/10.3847/1538-4357/aac267} {\bibfield  {journal}
  {\bibinfo  {journal} {Astrophys. J.}\ }\textbf {\bibinfo {volume} {860}},\
  \bibinfo {pages} {149} (\bibinfo {year} {2018})},\ \Eprint
  {https://arxiv.org/abs/1801.01923} {arXiv:1801.01923 [nucl-th]} \BibitemShut
  {NoStop}%
\bibitem [{\citenamefont {Read}\ \emph {et~al.}(2009)\citenamefont {Read},
  \citenamefont {Lackey}, \citenamefont {Owen},\ and\ \citenamefont
  {Friedman}}]{Read:2008iy}%
  \BibitemOpen
  \bibfield  {author} {\bibinfo {author} {\bibfnamefont {J.~S.}\ \bibnamefont
  {Read}}, \bibinfo {author} {\bibfnamefont {B.~D.}\ \bibnamefont {Lackey}},
  \bibinfo {author} {\bibfnamefont {B.~J.}\ \bibnamefont {Owen}},\ and\
  \bibinfo {author} {\bibfnamefont {J.~L.}\ \bibnamefont {Friedman}},\
  }\bibfield  {title} {\bibinfo {title} {{Constraints on a phenomenologically
  parameterized neutron-star equation of state}},\ }\href
  {https://doi.org/10.1103/PhysRevD.79.124032} {\bibfield  {journal} {\bibinfo
  {journal} {Phys. Rev. D}\ }\textbf {\bibinfo {volume} {79}},\ \bibinfo
  {pages} {124032} (\bibinfo {year} {2009})},\ \Eprint
  {https://arxiv.org/abs/0812.2163} {arXiv:0812.2163 [astro-ph]} \BibitemShut
  {NoStop}%
\bibitem [{\citenamefont {Bauswein}\ \emph {et~al.}(2010)\citenamefont
  {Bauswein}, \citenamefont {Janka},\ and\ \citenamefont
  {Oechslin}}]{Bauswein:2010dn}%
  \BibitemOpen
  \bibfield  {author} {\bibinfo {author} {\bibfnamefont {A.}~\bibnamefont
  {Bauswein}}, \bibinfo {author} {\bibfnamefont {H.~T.}\ \bibnamefont
  {Janka}},\ and\ \bibinfo {author} {\bibfnamefont {R.}~\bibnamefont
  {Oechslin}},\ }\bibfield  {title} {\bibinfo {title} {{Testing Approximations
  of Thermal Effects in Neutron Star Merger Simulations}},\ }\href
  {https://doi.org/10.1103/PhysRevD.82.084043} {\bibfield  {journal} {\bibinfo
  {journal} {Phys. Rev. D}\ }\textbf {\bibinfo {volume} {82}},\ \bibinfo
  {pages} {084043} (\bibinfo {year} {2010})},\ \Eprint
  {https://arxiv.org/abs/1006.3315} {arXiv:1006.3315 [astro-ph.SR]}
  \BibitemShut {NoStop}%
\bibitem [{\citenamefont {Tichy}(2009)}]{Tichy:2009yr}%
  \BibitemOpen
  \bibfield  {author} {\bibinfo {author} {\bibfnamefont {W.}~\bibnamefont
  {Tichy}},\ }\bibfield  {title} {\bibinfo {title} {{A New numerical method to
  construct binary neutron star initial data}},\ }\href
  {https://doi.org/10.1088/0264-9381/26/17/175018} {\bibfield  {journal}
  {\bibinfo  {journal} {Class. Quant. Grav.}\ }\textbf {\bibinfo {volume}
  {26}},\ \bibinfo {pages} {175018} (\bibinfo {year} {2009})},\ \Eprint
  {https://arxiv.org/abs/0908.0620} {arXiv:0908.0620 [gr-qc]} \BibitemShut
  {NoStop}%
\bibitem [{\citenamefont {Tichy}(2012)}]{Tichy:2012rp}%
  \BibitemOpen
  \bibfield  {author} {\bibinfo {author} {\bibfnamefont {W.}~\bibnamefont
  {Tichy}},\ }\bibfield  {title} {\bibinfo {title} {{Constructing
  quasi-equilibrium initial data for binary neutron stars with arbitrary
  spins}},\ }\href {https://doi.org/10.1103/PhysRevD.86.064024} {\bibfield
  {journal} {\bibinfo  {journal} {Phys. Rev. D}\ }\textbf {\bibinfo {volume}
  {86}},\ \bibinfo {pages} {064024} (\bibinfo {year} {2012})},\ \Eprint
  {https://arxiv.org/abs/1209.5336} {arXiv:1209.5336 [gr-qc]} \BibitemShut
  {NoStop}%
\bibitem [{\citenamefont {Tichy}\ \emph {et~al.}(2019)\citenamefont {Tichy},
  \citenamefont {Rashti}, \citenamefont {Dietrich}, \citenamefont {Dudi},\ and\
  \citenamefont {Br\"ugmann}}]{Tichy:2019ouu}%
  \BibitemOpen
  \bibfield  {author} {\bibinfo {author} {\bibfnamefont {W.}~\bibnamefont
  {Tichy}}, \bibinfo {author} {\bibfnamefont {A.}~\bibnamefont {Rashti}},
  \bibinfo {author} {\bibfnamefont {T.}~\bibnamefont {Dietrich}}, \bibinfo
  {author} {\bibfnamefont {R.}~\bibnamefont {Dudi}},\ and\ \bibinfo {author}
  {\bibfnamefont {B.}~\bibnamefont {Br\"ugmann}},\ }\bibfield  {title}
  {\bibinfo {title} {{Constructing binary neutron star initial data with high
  spins, high compactnesses, and high mass ratios}},\ }\href
  {https://doi.org/10.1103/PhysRevD.100.124046} {\bibfield  {journal} {\bibinfo
   {journal} {Phys. Rev. D}\ }\textbf {\bibinfo {volume} {100}},\ \bibinfo
  {pages} {124046} (\bibinfo {year} {2019})},\ \Eprint
  {https://arxiv.org/abs/1910.09690} {arXiv:1910.09690 [gr-qc]} \BibitemShut
  {NoStop}%
\bibitem [{\citenamefont {Wilson}\ and\ \citenamefont
  {Mathews}(1995)}]{Wilson:1995uh}%
  \BibitemOpen
  \bibfield  {author} {\bibinfo {author} {\bibfnamefont {J.~R.}\ \bibnamefont
  {Wilson}}\ and\ \bibinfo {author} {\bibfnamefont {G.~J.}\ \bibnamefont
  {Mathews}},\ }\bibfield  {title} {\bibinfo {title} {{Instabilities in Close
  Neutron Star Binaries}},\ }\href
  {https://doi.org/10.1103/PhysRevLett.75.4161} {\bibfield  {journal} {\bibinfo
   {journal} {Phys. Rev. Lett.}\ }\textbf {\bibinfo {volume} {75}},\ \bibinfo
  {pages} {4161} (\bibinfo {year} {1995})}\BibitemShut {NoStop}%
\bibitem [{\citenamefont {Wilson}\ \emph {et~al.}(1996)\citenamefont {Wilson},
  \citenamefont {Mathews},\ and\ \citenamefont {Marronetti}}]{Wilson:1996ty}%
  \BibitemOpen
  \bibfield  {author} {\bibinfo {author} {\bibfnamefont {J.~R.}\ \bibnamefont
  {Wilson}}, \bibinfo {author} {\bibfnamefont {G.~J.}\ \bibnamefont
  {Mathews}},\ and\ \bibinfo {author} {\bibfnamefont {P.}~\bibnamefont
  {Marronetti}},\ }\bibfield  {title} {\bibinfo {title} {{Relativistic
  numerical model for close neutron star binaries}},\ }\href
  {https://doi.org/10.1103/PhysRevD.54.1317} {\bibfield  {journal} {\bibinfo
  {journal} {Phys. Rev. D}\ }\textbf {\bibinfo {volume} {54}},\ \bibinfo
  {pages} {1317} (\bibinfo {year} {1996})},\ \Eprint
  {https://arxiv.org/abs/gr-qc/9601017} {arXiv:gr-qc/9601017} \BibitemShut
  {NoStop}%
\bibitem [{\citenamefont {York}(1999)}]{York:1998hy}%
  \BibitemOpen
  \bibfield  {author} {\bibinfo {author} {\bibfnamefont {J.~W.}\ \bibnamefont
  {York}, \bibfnamefont {Jr.}},\ }\bibfield  {title} {\bibinfo {title}
  {{Conformal 'thin sandwich' data for the initial-value problem}},\ }\href
  {https://doi.org/10.1103/PhysRevLett.82.1350} {\bibfield  {journal} {\bibinfo
   {journal} {Phys. Rev. Lett.}\ }\textbf {\bibinfo {volume} {82}},\ \bibinfo
  {pages} {1350} (\bibinfo {year} {1999})},\ \Eprint
  {https://arxiv.org/abs/gr-qc/9810051} {arXiv:gr-qc/9810051} \BibitemShut
  {NoStop}%
\bibitem [{\citenamefont {Bruegmann}\ \emph {et~al.}(2008)\citenamefont
  {Bruegmann}, \citenamefont {Gonzalez}, \citenamefont {Hannam}, \citenamefont
  {Husa}, \citenamefont {Sperhake},\ and\ \citenamefont
  {Tichy}}]{Bruegmann:2006ulg}%
  \BibitemOpen
  \bibfield  {author} {\bibinfo {author} {\bibfnamefont {B.}~\bibnamefont
  {Bruegmann}}, \bibinfo {author} {\bibfnamefont {J.~A.}\ \bibnamefont
  {Gonzalez}}, \bibinfo {author} {\bibfnamefont {M.}~\bibnamefont {Hannam}},
  \bibinfo {author} {\bibfnamefont {S.}~\bibnamefont {Husa}}, \bibinfo {author}
  {\bibfnamefont {U.}~\bibnamefont {Sperhake}},\ and\ \bibinfo {author}
  {\bibfnamefont {W.}~\bibnamefont {Tichy}},\ }\bibfield  {title} {\bibinfo
  {title} {{Calibration of Moving Puncture Simulations}},\ }\href
  {https://doi.org/10.1103/PhysRevD.77.024027} {\bibfield  {journal} {\bibinfo
  {journal} {Phys. Rev. D}\ }\textbf {\bibinfo {volume} {77}},\ \bibinfo
  {pages} {024027} (\bibinfo {year} {2008})},\ \Eprint
  {https://arxiv.org/abs/gr-qc/0610128} {arXiv:gr-qc/0610128} \BibitemShut
  {NoStop}%
\bibitem [{\citenamefont {Thierfelder}\ \emph {et~al.}(2011)\citenamefont
  {Thierfelder}, \citenamefont {Bernuzzi},\ and\ \citenamefont
  {Bruegmann}}]{Thierfelder:2011yi}%
  \BibitemOpen
  \bibfield  {author} {\bibinfo {author} {\bibfnamefont {M.}~\bibnamefont
  {Thierfelder}}, \bibinfo {author} {\bibfnamefont {S.}~\bibnamefont
  {Bernuzzi}},\ and\ \bibinfo {author} {\bibfnamefont {B.}~\bibnamefont
  {Bruegmann}},\ }\bibfield  {title} {\bibinfo {title} {{Numerical relativity
  simulations of binary neutron stars}},\ }\href
  {https://doi.org/10.1103/PhysRevD.84.044012} {\bibfield  {journal} {\bibinfo
  {journal} {Phys. Rev. D}\ }\textbf {\bibinfo {volume} {84}},\ \bibinfo
  {pages} {044012} (\bibinfo {year} {2011})},\ \Eprint
  {https://arxiv.org/abs/1104.4751} {arXiv:1104.4751 [gr-qc]} \BibitemShut
  {NoStop}%
\bibitem [{\citenamefont {Dietrich}\ \emph
  {et~al.}(2015{\natexlab{b}})\citenamefont {Dietrich}, \citenamefont
  {Bernuzzi}, \citenamefont {Ujevic},\ and\ \citenamefont
  {Br\"ugmann}}]{Dietrich:2015iva}%
  \BibitemOpen
  \bibfield  {author} {\bibinfo {author} {\bibfnamefont {T.}~\bibnamefont
  {Dietrich}}, \bibinfo {author} {\bibfnamefont {S.}~\bibnamefont {Bernuzzi}},
  \bibinfo {author} {\bibfnamefont {M.}~\bibnamefont {Ujevic}},\ and\ \bibinfo
  {author} {\bibfnamefont {B.}~\bibnamefont {Br\"ugmann}},\ }\bibfield  {title}
  {\bibinfo {title} {{Numerical relativity simulations of neutron star merger
  remnants using conservative mesh refinement}},\ }\href
  {https://doi.org/10.1103/PhysRevD.91.124041} {\bibfield  {journal} {\bibinfo
  {journal} {Phys. Rev. D}\ }\textbf {\bibinfo {volume} {91}},\ \bibinfo
  {pages} {124041} (\bibinfo {year} {2015}{\natexlab{b}})},\ \Eprint
  {https://arxiv.org/abs/1504.01266} {arXiv:1504.01266 [gr-qc]} \BibitemShut
  {NoStop}%
\bibitem [{\citenamefont {Berger}\ and\ \citenamefont
  {Oliger}(1984)}]{Berger:1984zza}%
  \BibitemOpen
  \bibfield  {author} {\bibinfo {author} {\bibfnamefont {M.~J.}\ \bibnamefont
  {Berger}}\ and\ \bibinfo {author} {\bibfnamefont {J.}~\bibnamefont
  {Oliger}},\ }\bibfield  {title} {\bibinfo {title} {{Adaptive Mesh Refinement
  for Hyperbolic Partial Differential Equations}},\ }\href
  {https://doi.org/10.1016/0021-9991(84)90073-1} {\bibfield  {journal}
  {\bibinfo  {journal} {J. Comput. Phys.}\ }\textbf {\bibinfo {volume} {53}},\
  \bibinfo {pages} {484} (\bibinfo {year} {1984})}\BibitemShut {NoStop}%
\bibitem [{\citenamefont {Bernuzzi}\ \emph {et~al.}(2012)\citenamefont
  {Bernuzzi}, \citenamefont {Nagar}, \citenamefont {Thierfelder},\ and\
  \citenamefont {Brugmann}}]{Bernuzzi:2012ci}%
  \BibitemOpen
  \bibfield  {author} {\bibinfo {author} {\bibfnamefont {S.}~\bibnamefont
  {Bernuzzi}}, \bibinfo {author} {\bibfnamefont {A.}~\bibnamefont {Nagar}},
  \bibinfo {author} {\bibfnamefont {M.}~\bibnamefont {Thierfelder}},\ and\
  \bibinfo {author} {\bibfnamefont {B.}~\bibnamefont {Brugmann}},\ }\bibfield
  {title} {\bibinfo {title} {{Tidal effects in binary neutron star
  coalescence}},\ }\href {https://doi.org/10.1103/PhysRevD.86.044030}
  {\bibfield  {journal} {\bibinfo  {journal} {Phys. Rev. D}\ }\textbf {\bibinfo
  {volume} {86}},\ \bibinfo {pages} {044030} (\bibinfo {year} {2012})},\
  \Eprint {https://arxiv.org/abs/1205.3403} {arXiv:1205.3403 [gr-qc]}
  \BibitemShut {NoStop}%
\bibitem [{\citenamefont {Newman}\ and\ \citenamefont
  {Penrose}(1962)}]{Newman:1961qr}%
  \BibitemOpen
  \bibfield  {author} {\bibinfo {author} {\bibfnamefont {E.}~\bibnamefont
  {Newman}}\ and\ \bibinfo {author} {\bibfnamefont {R.}~\bibnamefont
  {Penrose}},\ }\bibfield  {title} {\bibinfo {title} {{An Approach to
  gravitational radiation by a method of spin coefficients}},\ }\href
  {https://doi.org/10.1063/1.1724257} {\bibfield  {journal} {\bibinfo
  {journal} {J. Math. Phys.}\ }\textbf {\bibinfo {volume} {3}},\ \bibinfo
  {pages} {566} (\bibinfo {year} {1962})}\BibitemShut {NoStop}%
\bibitem [{\citenamefont {Reisswig}\ and\ \citenamefont
  {Pollney}(2011)}]{Reisswig:2010di}%
  \BibitemOpen
  \bibfield  {author} {\bibinfo {author} {\bibfnamefont {C.}~\bibnamefont
  {Reisswig}}\ and\ \bibinfo {author} {\bibfnamefont {D.}~\bibnamefont
  {Pollney}},\ }\bibfield  {title} {\bibinfo {title} {{Notes on the integration
  of numerical relativity waveforms}},\ }\href
  {https://doi.org/10.1088/0264-9381/28/19/195015} {\bibfield  {journal}
  {\bibinfo  {journal} {Class. Quant. Grav.}\ }\textbf {\bibinfo {volume}
  {28}},\ \bibinfo {pages} {195015} (\bibinfo {year} {2011})},\ \Eprint
  {https://arxiv.org/abs/1006.1632} {arXiv:1006.1632 [gr-qc]} \BibitemShut
  {NoStop}%
\bibitem [{\citenamefont {Ujevic}\ \emph {et~al.}(2022)\citenamefont {Ujevic},
  \citenamefont {Rashti}, \citenamefont {Gieg}, \citenamefont {Tichy},\ and\
  \citenamefont {Dietrich}}]{Ujevic:2022qle}%
  \BibitemOpen
  \bibfield  {author} {\bibinfo {author} {\bibfnamefont {M.}~\bibnamefont
  {Ujevic}}, \bibinfo {author} {\bibfnamefont {A.}~\bibnamefont {Rashti}},
  \bibinfo {author} {\bibfnamefont {H.}~\bibnamefont {Gieg}}, \bibinfo {author}
  {\bibfnamefont {W.}~\bibnamefont {Tichy}},\ and\ \bibinfo {author}
  {\bibfnamefont {T.}~\bibnamefont {Dietrich}},\ }\bibfield  {title} {\bibinfo
  {title} {{High-accuracy high-mass-ratio simulations for binary neutron stars
  and their comparison to existing waveform models}},\ }\href
  {https://doi.org/10.1103/PhysRevD.106.023029} {\bibfield  {journal} {\bibinfo
   {journal} {Phys. Rev. D}\ }\textbf {\bibinfo {volume} {106}},\ \bibinfo
  {pages} {023029} (\bibinfo {year} {2022})},\ \Eprint
  {https://arxiv.org/abs/2202.09343} {arXiv:2202.09343 [gr-qc]} \BibitemShut
  {NoStop}%
\bibitem [{\citenamefont {Dietrich}\ \emph {et~al.}(2017)\citenamefont
  {Dietrich}, \citenamefont {Ujevic}, \citenamefont {Tichy}, \citenamefont
  {Bernuzzi},\ and\ \citenamefont {Bruegmann}}]{Dietrich:2016hky}%
  \BibitemOpen
  \bibfield  {author} {\bibinfo {author} {\bibfnamefont {T.}~\bibnamefont
  {Dietrich}}, \bibinfo {author} {\bibfnamefont {M.}~\bibnamefont {Ujevic}},
  \bibinfo {author} {\bibfnamefont {W.}~\bibnamefont {Tichy}}, \bibinfo
  {author} {\bibfnamefont {S.}~\bibnamefont {Bernuzzi}},\ and\ \bibinfo
  {author} {\bibfnamefont {B.}~\bibnamefont {Bruegmann}},\ }\bibfield  {title}
  {\bibinfo {title} {{Gravitational waves and mass ejecta from binary neutron
  star mergers: Effect of the mass-ratio}},\ }\href
  {https://doi.org/10.1103/PhysRevD.95.024029} {\bibfield  {journal} {\bibinfo
  {journal} {Phys. Rev. D}\ }\textbf {\bibinfo {volume} {95}},\ \bibinfo
  {pages} {024029} (\bibinfo {year} {2017})},\ \Eprint
  {https://arxiv.org/abs/1607.06636} {arXiv:1607.06636 [gr-qc]} \BibitemShut
  {NoStop}%
\end{thebibliography}%

\appendix

\section{Waveform Analysis} \label{app:waves}

\begin{figure}[t]
  \includegraphics[width=0.52\textwidth]{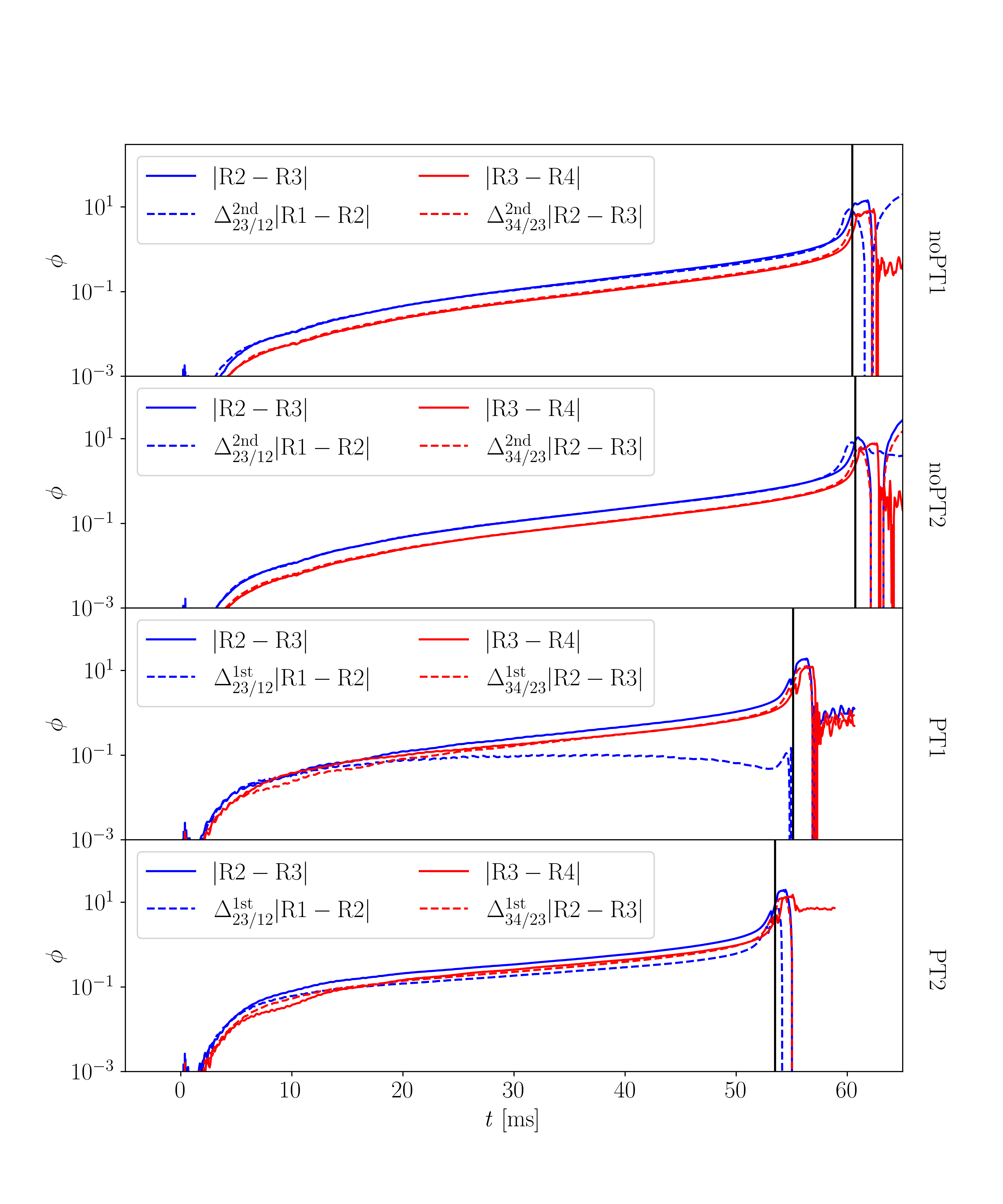}
\caption{Convergence of the (2,2)-mode for all the setups. We show the GW phase differences between consecutive resolutions and the rescaled phase differences assuming second order convergence for the noPT cases and first order for the PT cases. Black vertical line marks the merger time.} 

\label{fig:convergence}
\end{figure}

To compute GWs, we follow the formalism of Ref.~\cite{Newman:1961qr} and extract the Newman-Penrose invariant $\Psi_4$ on a sphere of radius $r \simeq 1200$~km. We then reconstruct the strain $h$ using the scheme depicted in \cite{Reisswig:2010di}.

As a first check, we perform a self convergence analysis of the GW phase of the (2,2)-mode for every configuration using all four resolutions; cf.~Fig.~\ref{fig:convergence}. We obtain second order convergence for the noPT configurations, in agreement with our previous findings~\cite{Bernuzzi:2016pie,Ujevic:2022qle}, and first order for the PT ones~\cite{Gieg:2019yzq}. 

Phase errors due to finite resolution effects are computed as 
\begin{equation}
    \Delta \phi^{}_{\rm res} = |\phi^{}_{\rm Ric} - \phi^{}_{\rm R4}|,
    \label{eq:delta_res}
\end{equation}

\noindent where $\phi^{}_{\rm Ric}$ is the phase obtained by Richardson extrapolation \cite{Bernuzzi:2016pie} from R3 and R4.

Since the presence of a phase transition also complicates the computational of the initial data -- due to the presence of non-smooth fields inside the star -- we find generally higher constraint violations, when we employ our standard SGRID resolution. To investigate this effect, we perform one simulation for which we increase the SGRID-resolution from typically $26^3$ points per individual grid towards $38^3$ points; cf.~Fig.~2 of~\cite{Tichy:2019ouu}. We evolve this setup with resolution R3 and compute the difference for the different SGRID resolutions:
\begin{equation}
    \Delta \phi^{}_{\rm ID} = |\phi^{}_{\rm ID} - \phi^{}_{\rm R3}|\,,
        \label{eq:delta_ID}
\end{equation}
with $\phi^{}_{\rm ID}$ being the phase in the simulation with high resolution ID. This difference is added in quadrature to the PT1 and noPT1 finite resolution error shown in Fig.~\ref{fig:GW1}: 
\begin{equation}
    \Delta \phi^{}_{\rm tot} = \sqrt{ \Delta \phi_{\rm res}^2 + \Delta \phi_{\rm ID}^2}\,.
    \label{eq:delta_tot}
\end{equation}

\begin{figure}[t!]
\centering
  \centering
  \includegraphics[width=\columnwidth]{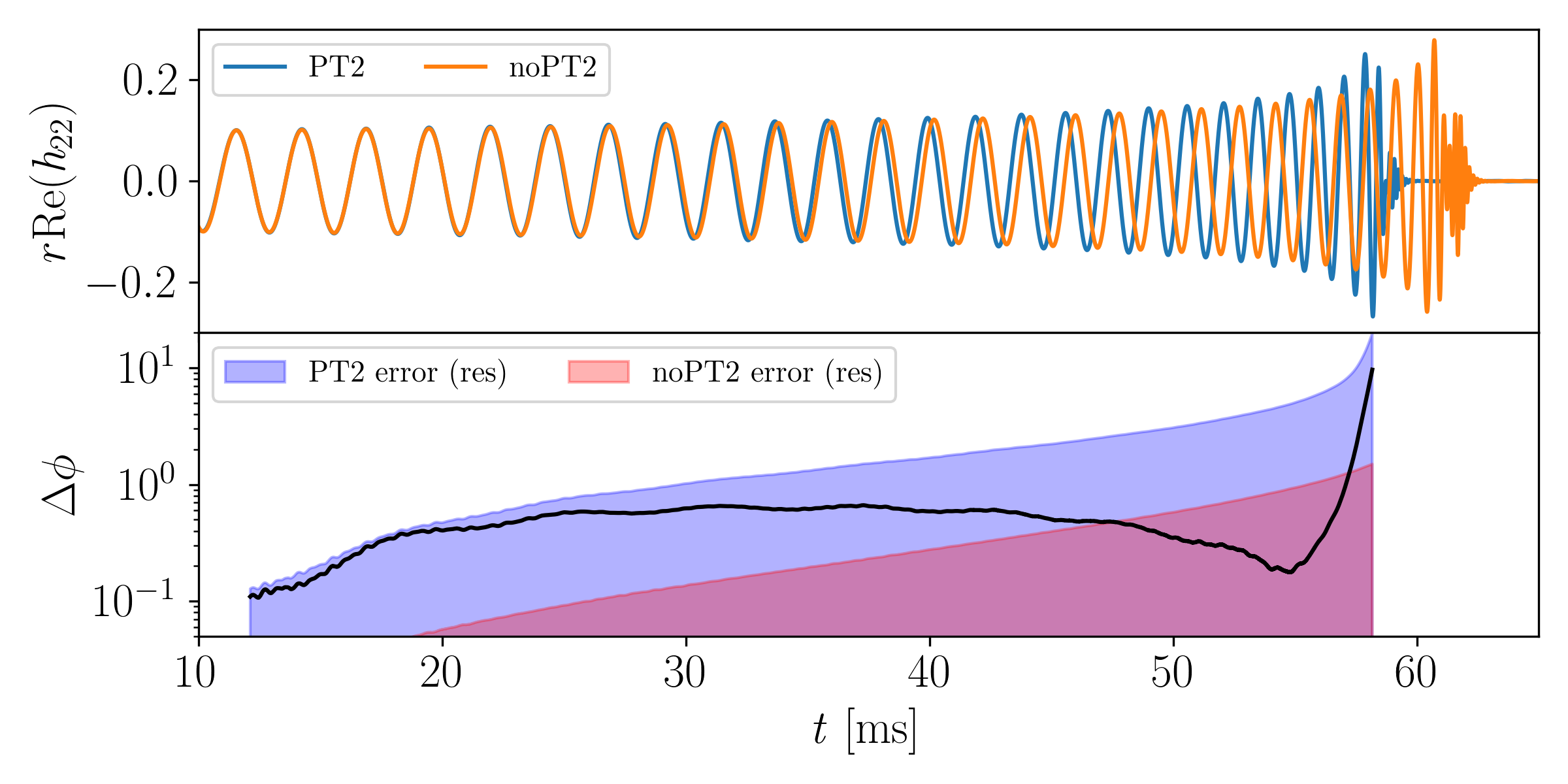}
\caption{Top panel: (2,2)-mode of the highest resolution waveform for PT2 (blue) and noPT2 (orange) scenarios. {Bottom panel:} The black line shows the GW phase difference between the Richardson extrapolated waveform for PT2 and noPT2. Shaded areas show the error on the GW $\phi$ for each system computed with Eq. (\ref{eq:delta_res}). In contrast to Fig.~\ref{fig:GW1} we do not incorporate an additional phase difference based on simulations with different ID resolutions.}  
\label{fig:GW2}
\end{figure}

Similar to the results shown for PT1 and noPT1, we also present the dephasing between PT2 and noPT2 in Fig.~\ref{fig:GW2} and find, as before, that the phase difference is smaller than the uncertainty of the simulations. 
This is mostly due to the large phase error connected to the PT simulations.  

\section{PT2 and noPT2 Simulations} \label{app:extra}

\begin{figure*}[t]
\centering
        \hspace{0.3cm}\includegraphics[width=1.92\columnwidth]{cbar_noPT_rho.pdf}\\
        \vspace{0.3cm}
  \includegraphics[width=0.48\columnwidth]{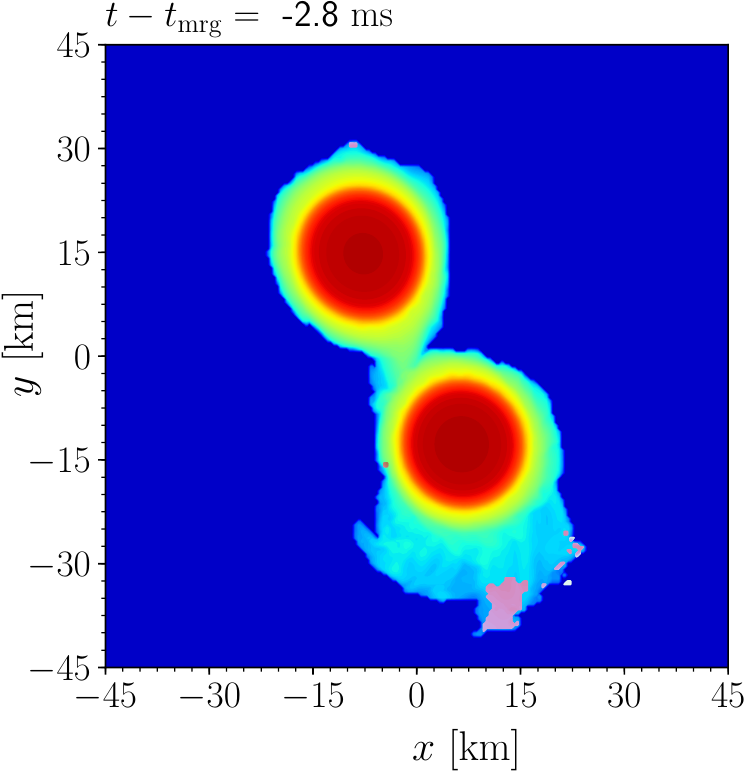} 
    \includegraphics[width=0.48\columnwidth]{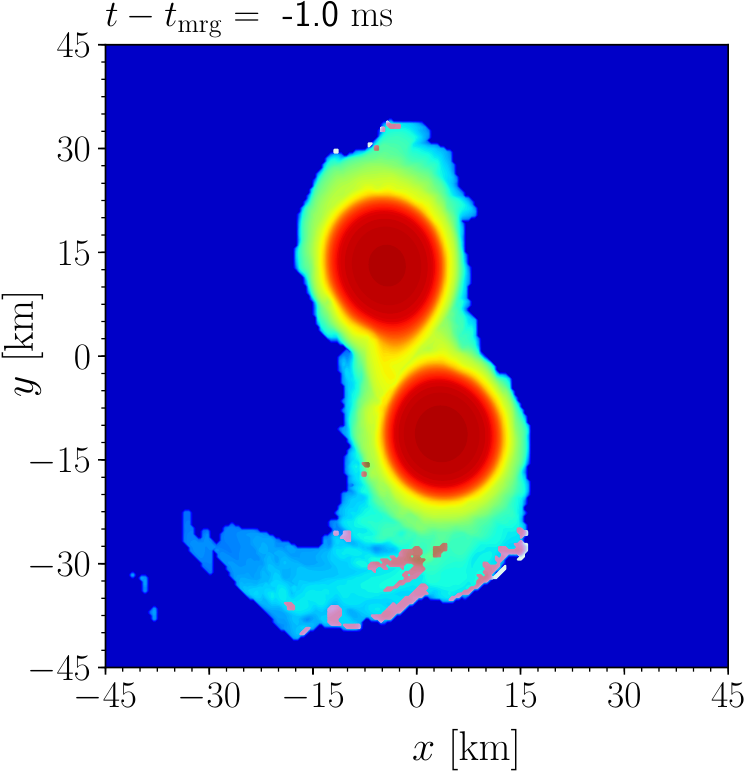} 
      \includegraphics[width=0.48\columnwidth]{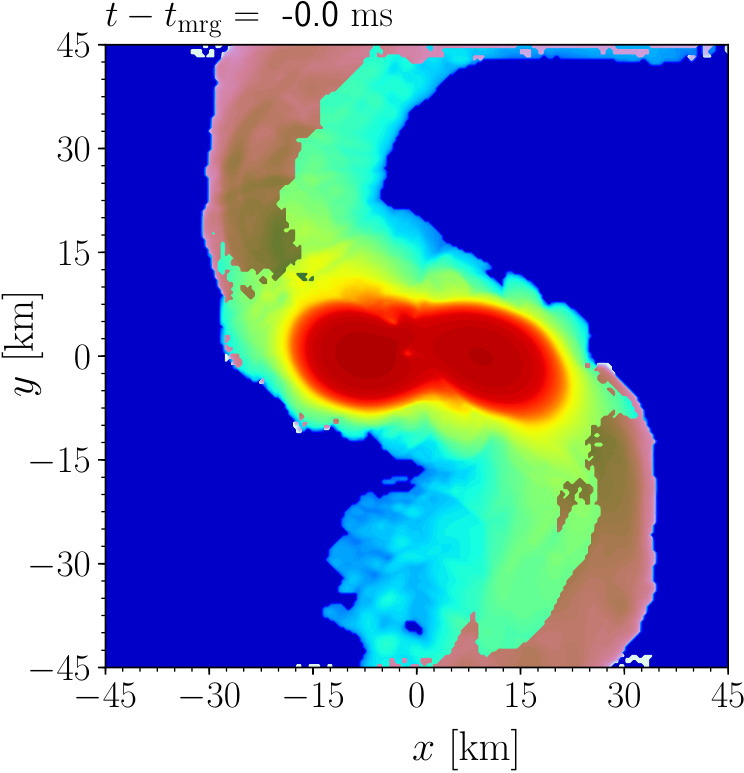} 
        \includegraphics[width=0.48\columnwidth]{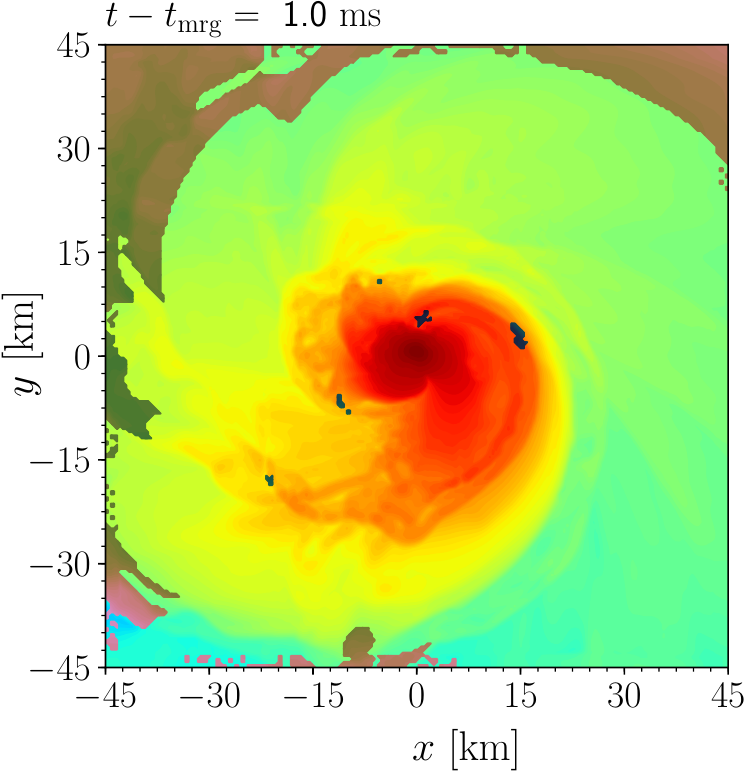} \\
	\hspace{0.05cm}\includegraphics[width=0.48\columnwidth, height=0.4\columnwidth]{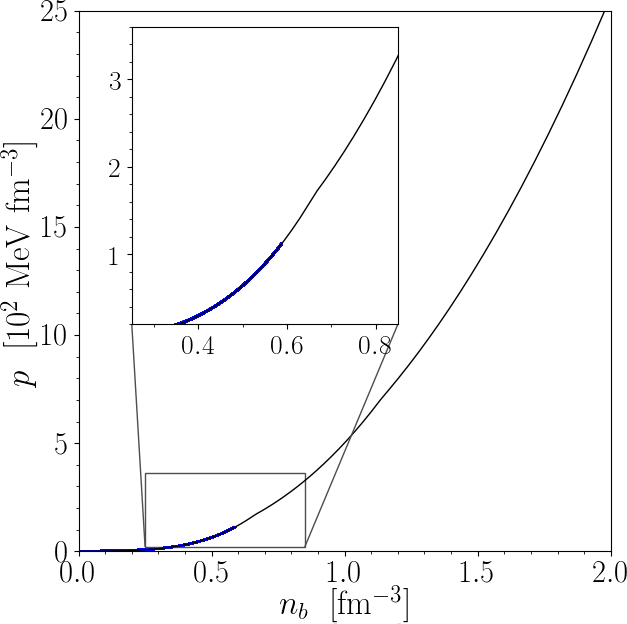}
 \includegraphics[width=0.48\columnwidth, height=0.4\columnwidth]{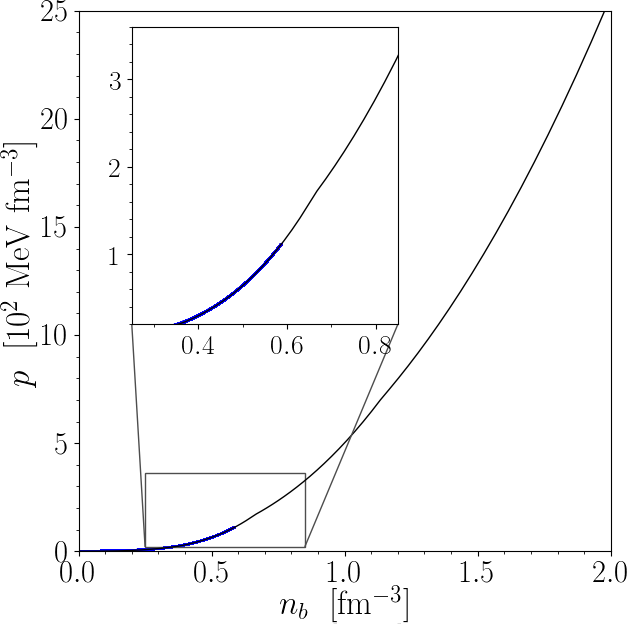}
      \includegraphics[width=0.48\columnwidth, height=0.4\columnwidth]{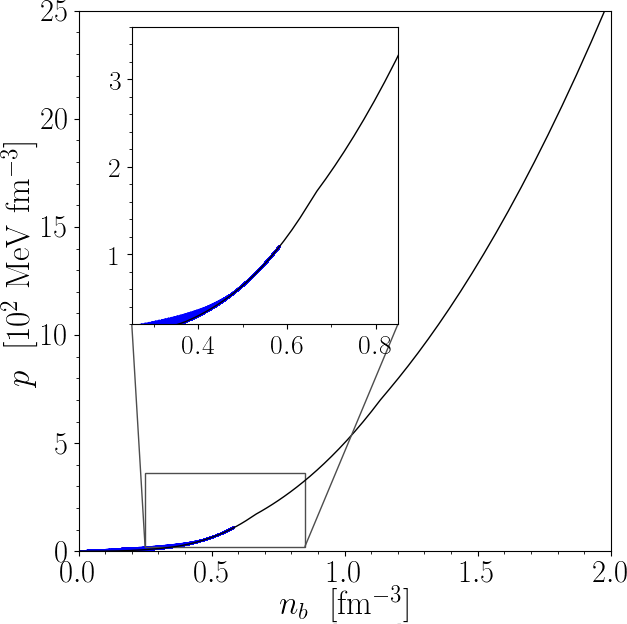}
        \includegraphics[width=0.48\columnwidth, height=0.4\columnwidth]{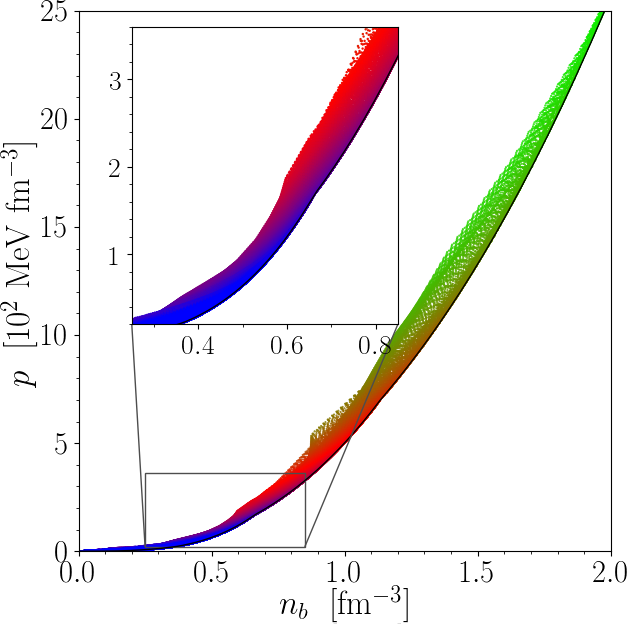}\\
                \hspace{0.3cm}\includegraphics[width=1.92\columnwidth]{colorbar_epsilon.pdf}
\caption{Snapshots of the baryon number density on the equatorial plane (upper panels) and  the respective baryon number density-pressure phase diagram (lower panels), together with the zero-temperature EOS (black line) and color-coded specific internal energy, are depicted for the noPT2 setup employing the highest resolution.
Hadronic matter is represented by the upper left colorbar and matter identified as ejecta by the upper right colorbar.} 
\label{fig:2d_noPT02}
\end{figure*}

\begin{figure*}[t]
\centering
        \hspace{0.3cm}\includegraphics[width=1.92\columnwidth]{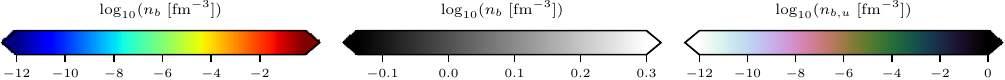}\\
        \vspace{0.3cm}
  \includegraphics[width=0.48\columnwidth]{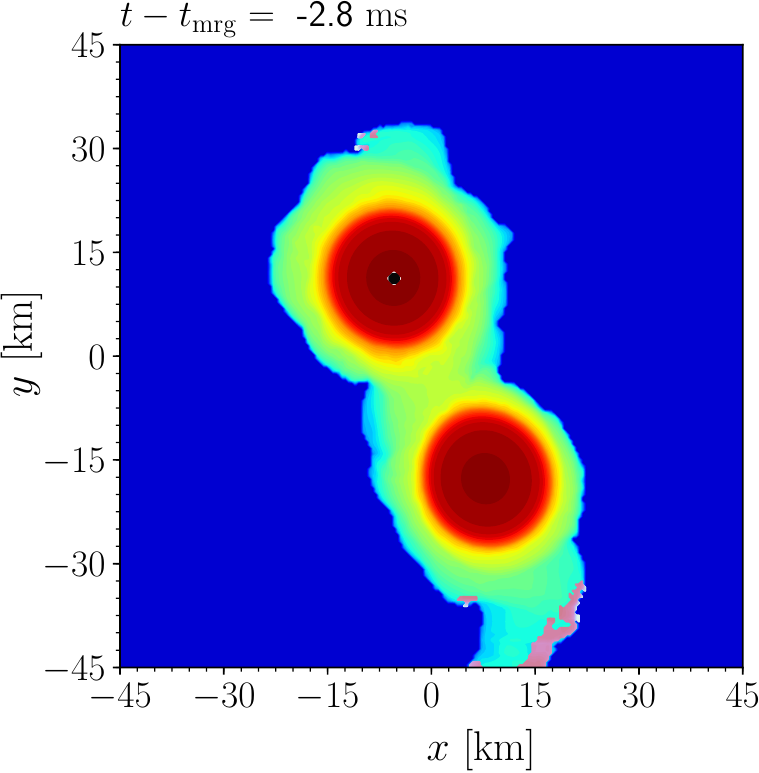} 
    \includegraphics[width=0.48\columnwidth]{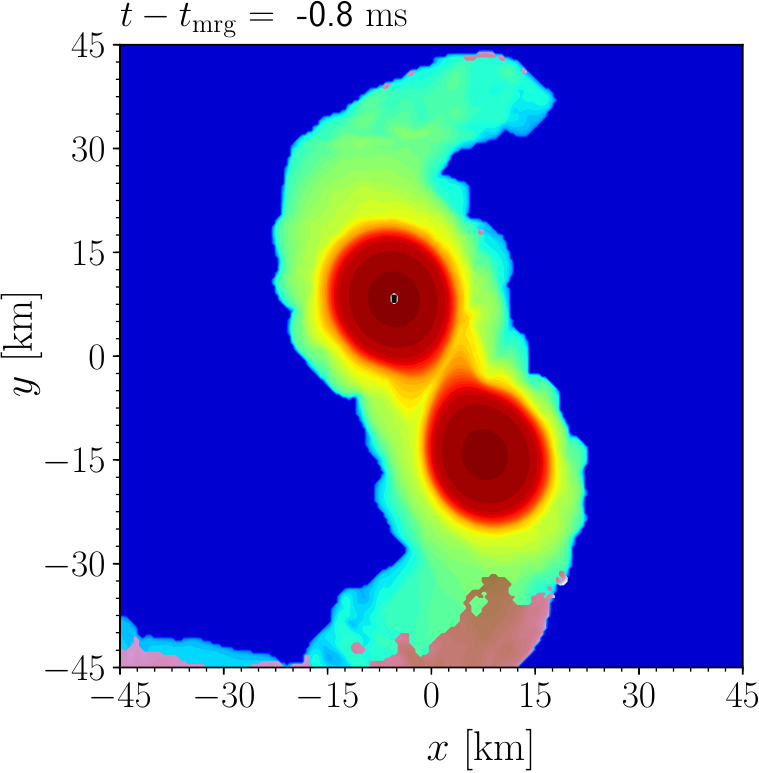} 
      \includegraphics[width=0.48\columnwidth]{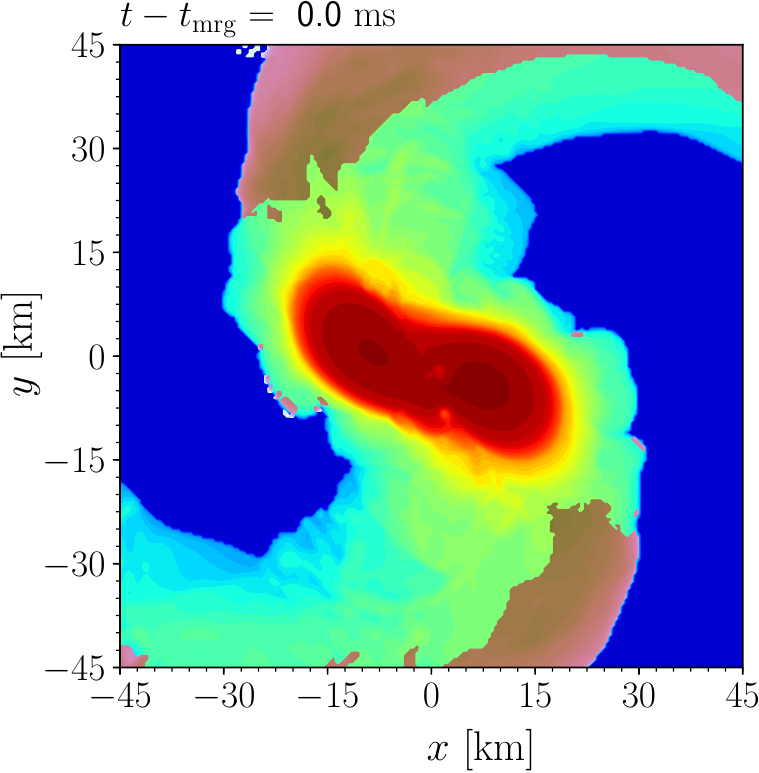} 
        \includegraphics[width=0.48\columnwidth]{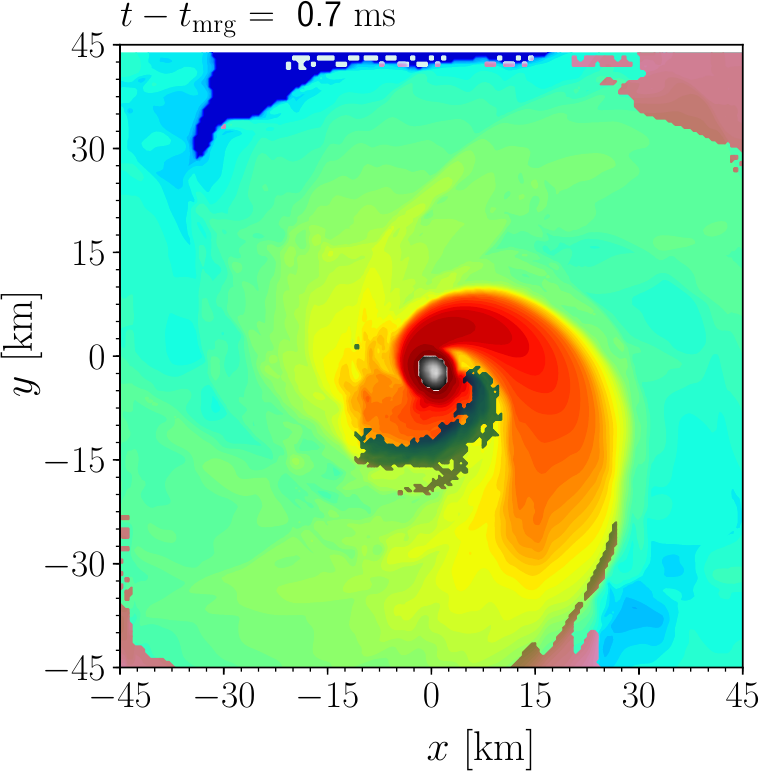} \\
	\hspace{0.05cm}\includegraphics[width=0.48\columnwidth, height=0.4\columnwidth]{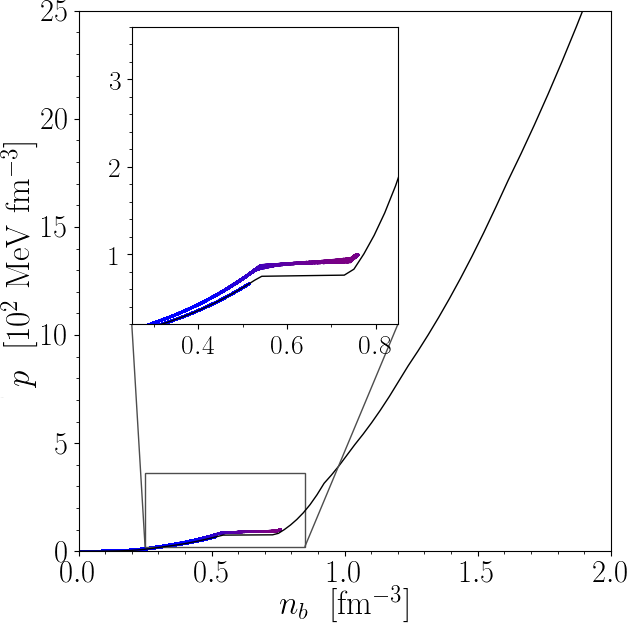}
 \includegraphics[width=0.48\columnwidth, height=0.4\columnwidth]{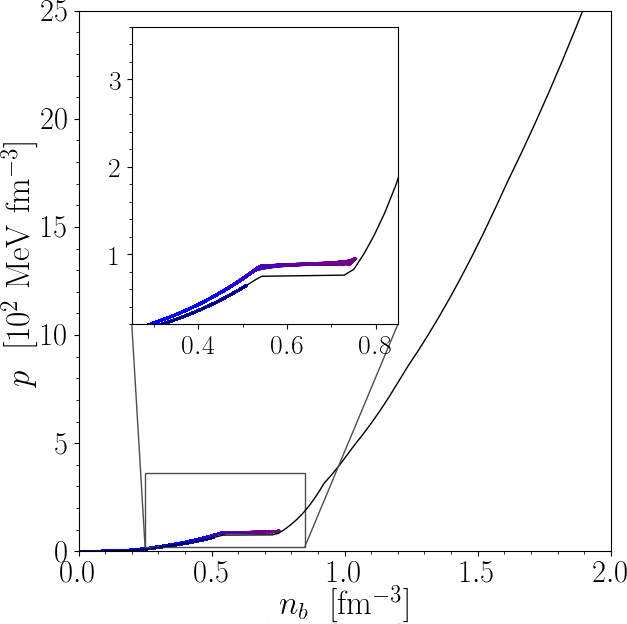}
      \includegraphics[width=0.48\columnwidth, height=0.4\columnwidth]{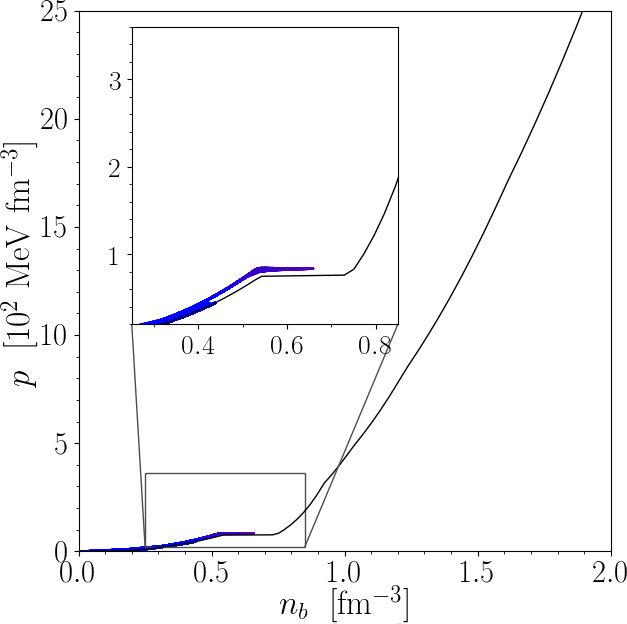}
        \includegraphics[width=0.48\columnwidth, height=0.4\columnwidth]{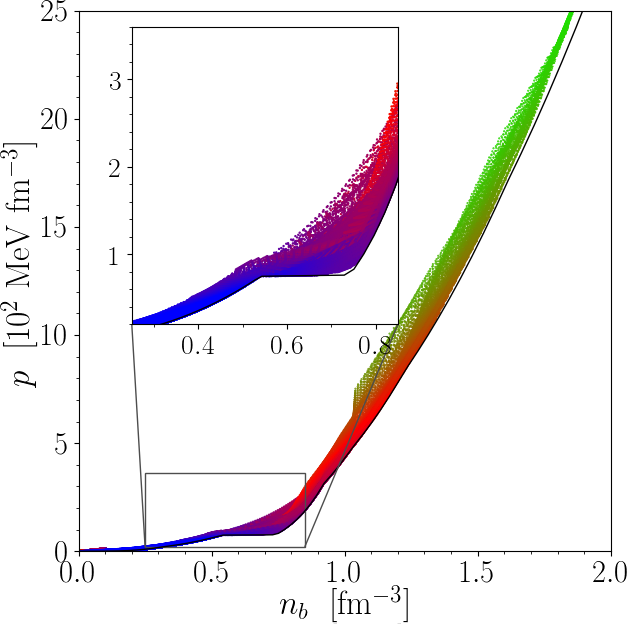}\\
                \hspace{0.3cm}\includegraphics[width=1.92\columnwidth]{colorbar_epsilon.pdf}
\caption{Snapshots of the baryon number density on the equatorial plane (upper panels) and the respective baryon number density-pressure phase diagram (lower panels) for the PT2 case are depicted employing the highest resolution. Similarly to Fig.~\ref{fig:2D_PT01}, we represent hadronic matter in the upper left colorbar, the quark matter using the upper middle colorbar and the ejecta using the colorbar on the upper right. 
The lower colorbar refers to the specific internal energy. 
         \textit{Upper panels, first column}: at this instance we notice that the inner core of one of the stars contains quarks at $\log_{10}(n_b~[{\rm fm^{-3}}]) \sim - 0.1$. 
         \textit{Upper panels, second column}: the densest portions of the stars are found tidally deformed. The inner core containing quarks is smaller and less dense with $\log_{10}(n_b~[{\rm fm^{-3}}]) < -0.1$. 
         \textit{Upper panels, third column}: moment of merger, where we note the complete absence of quarks in the coalescing material. 
         \textit{Upper panels, fourth column}: in less than $1~{\rm ms}$ after the merger a sizeable quark-matter core is formed, which then quickly collapses to a BH. 
         \textit{Lower panels}: The zero-temperature PT2 EOS is depicted (black line) along with the color-coded specific internal energies for the respective snapshots; similarly to the PT1 case, we see the decrease of the quarks content within the star towards the merger and its subsequent increase rapidly after the merger.} 
\label{fig:2D_PT02}
\end{figure*}

In Figs.~\ref{fig:2d_noPT02} and~\ref{fig:2D_PT02} we present snapshots of the baryon number density in the orbital plane (upper panels) and the corresponding $n_b$-$p$-$\epsilon$-diagrams (lower panels) for the noPT2 and PT2 highest resolution simulation. 
Overall, the conclusions drawn from the PT1 setup regarding the reverse phase transition are also verified on the unequal-mass PT2 setup, whose EOS is also different than that of the PT1 setup. 
Likewise, we notice the vanishing of the quark-matter core of the more massive star soon before the merger and its reappearance rapidly after the merger. 
In both noPT2 and PT2 cases, the remnant collapses to a black hole for all resolutions. 
Similarly to the lower panels (fourth column) of Figs.~\ref{fig:2D_noPT1} and~\ref{fig:2D_PT01}, we see that larger pressures are achieved by the core of the noPT2 setup due to thermal contributions. For the mass-ratio $q=1.17$ considered, shock driven outflows dominate the ejection for soft EoSs~\cite{Dietrich:2016hky}, this reasonably suggests that its larger ejecta can be explained by a stronger core bounce during merger. Furthermore, the difference between the ejecta masses of the noPT2 and PT2 cases when compared to the noPT1 and PT1 cases, can be explained as a consequence of larger tidal tails close to the merger. 

Finally, it is worth pointing out that in the lower panels of Fig.~\ref{fig:2D_PT02} (first, second and third columns), one can see two branches in the $n_b$-$p$-$\epsilon$ diagram. The curve that is closer to the cold EOS refers to the hadronic star, while the curve that lies well above the cold EOS refers to points within the more massive star containing a quark-matter core. This shows, and we verified, that the thermal pressure is larger in this star, which is probably caused by shock heating of the matter around the discontinuity surface that separates the phases. 
This effect appears also in the PT1 simulations, although it is not visible in the $n_b$-$p$-$\epsilon$ diagrams due to the mass symmetry of the binary.

\section{Piecewise-polytropic representation of the employed Equations of State} \label{app:pwp}

In Tabs.~\ref{tab:PT} and \ref{tab:noPT} we present the piecewise-polytrope parameters of the EOSs used during the article. 

\begin{table*}[t]
\caption{EOS PT1 and EOS PT2. Piecewise polytrope of the EOSs with phase transition used in the simulations. The units are chosen so that the rest-mass density $\rho$ is in g/cm$^3$, $\Gamma$ is dimensionless and $K$ is such that the pressure $p$ is in g/cm$^3$. The $i$-th row refers to the polytrope $p = K_i \rho^{\Gamma_i}$ , for $\rho_i \leq \rho \leq \rho_{i + 1}$.} \label{tab:PT}
\begin{tabular}{ccc}
\toprule
\multicolumn{3}{c}{EOS PT1} \\ \hline
$\rho$ & $K$ & $\Gamma$ \\ \hline
0.0000 & $1.0830 \times 10^{-9}$ & 1.7392 \\
$6.5038 \times 10^5$ & $4.4916 \times 10^{-8}$ & 1.4609 \\
$7.7322 \times 10^7$ & $8.7621 \times 10^{-7}$ & 1.2974 \\
$2.2207 \times 10^{10}$ & $2.3406 \times 10^{-6}$ & 1.2561 \\
$3.3450 \times 10^{11}$ & $1.1756 \times 10^1$ & 0.6747 \\
$2.6359 \times 10^{12}$ & $3.0547 \times 10^{-7}$ & 1.2854 \\
$1.3459 \times 10^{13}$ & $1.9894 \times 10^{-9}$ & 1.4519 \\
$5.9323 \times 10^{13}$ & $6.5167 \times 10^{-13}$ & 1.7049 \\
$1.4646 \times 10^{14}$ & $3.5779 \times 10^{-22}$ & 2.3586 \\
$2.5283 \times 10^{14}$ & $1.6287 \times 10^{-34}$ & 3.2155 \\
$7.4119 \times 10^{14}$ & $3.3732 \times 10^{12}$ & 0.1007 \\
$1.0555 \times 10^{15}$ & $5.0936 \times 10^{-102}$ & 7.6770 \\
$1.1872 \times 10^{15}$ & $3.2385 \times 10^{-49}$ & 4.1742 \\
$1.4553 \times 10^{15}$ & $1.3742 \times 10^{-27}$ & 2.7478 \\
$2.0038 \times 10^{15}$ & $2.4168 \times 10^{-18}$  & 2.1436 \\ \hline \hline
\end{tabular}
\quad
\begin{tabular}{ccc}
\toprule \multicolumn{3}{c}{EOS PT2} \\ \hline
$\rho$ & $K$ & $\Gamma$ \\ \hline
0.0000 & $1.0830 \times 10^{-9}$ & 1.7392 \\
$6.5038 \times 10^5$ & $4.4916 \times 10^{-8}$ & 1.4609 \\
$7.9642 \times 10^7$ & $9.1355 \times 10^{-7}$ & 1.2953 \\
$3.9667 \times 10^{10}$ & $3.5047 \times 10^{-6}$ & 1.2402 \\
$3.0603 \times 10^{11}$ & 4.2083 & 0.7109 \\
$3.0664 \times 10^{12}$ & $2.3870 \times 10^{-8}$ & 1.3713 \\
$4.3127 \times 10^{13}$ & $2.5616 \times 10^{-12}$ & 1.6625 \\
$1.4409 \times 10^{14}$ & $1.8923 \times 10^{-22}$ & 2.3780 \\
$2.7367 \times 10^{14}$ & $1.8769 \times 10^{-37}$  & 3.4173 \\
$4.9468 \times 10^{14}$ & $1.7564 \times 10^{-23}$ & 2.4665 \\
$8.9714 \times 10^{14}$ & $1.7878 \times 10^{13}$ & 0.0584 \\
$1.2303 \times 10^{15}$ & $1.4283 \times 10^{-84}$ & 6.4929 \\
$1.5285 \times 10^{15}$ & $5.0642 \times 10^{-45}$ & 3.8883 \\
$1.7041 \times 10^{15}$ & $5.5642 \times 10^{-34}$ & 3.1634 \\
$2.0354 \times 10^{15}$ & $1.1130 \times 10^{-25}$ & 2.6212 \\
$2.6666 \times 10^{15}$ & $1.5627 \times 10^{-21}$ & 2.3523 \\ \hline \hline
\end{tabular}
\end{table*}

\begin{table*}[t]
\caption{EOS noPT1 and EOS noPT2. Piecewise polytrope of the EOSs without phase transition used in the simulations. The units are chosen so that the rest-mass density $\rho$ is in g/cm$^3$, $\Gamma$ is dimensionless and $K$ is such that the pressure $p$ is in g/cm$^3$. The $i$-th row refers to the polytrope $p = K_i \rho^{\Gamma_i}$ , for $\rho_i \leq \rho \leq \rho_{i + 1}$.} \label{tab:noPT}
\begin{tabular}{ccc}
\toprule
\multicolumn{3}{c}{EOS noPT1} \\ \hline
$\rho$ & $K$ & $\Gamma$ \\ \hline
0.0000 & $1.2847 \times 10^{-9}$ & 1.7209 \\
$1.2372 \times 10^6$ & $8.2700 \times 10^{-8}$ & 1.4241 \\
$1.6870 \times 10^8$ & $1.2124 \times 10^{-6}$ & 1.2823 \\ 
$2.7977 \times 10^{11}$ & 4.2083 & 0.7109 \\ 
$3.4214 \times 10^{12}$ & $4.9666 \times 10^{-9}$ & 1.4232 \\
$7.8043 \times 10^{13}$ & $5.5345 \times 10^{-16}$ & 1.9237 \\
$2.0594 \times 10^{14}$ & $1.4908 \times 10^{-28}$ & 2.8019 \\
$5.9849 \times 10^{14}$ & $4.2137 \times 10^{-38}$ & 3.4481 \\
$1.0962 \times 10^{15}$ & $8.8935 \times 10^{-26}$ & 2.6286 \\
$1.8879 \times 10^{15}$ & $1.7047 \times 10^{-20}$ & 2.2828 \\ \hline \hline
\end{tabular}
\quad
\begin{tabular}{ccc}
\toprule \multicolumn{3}{c}{EOS noPT2} \\ \hline
$\rho$ & $K$ & $\Gamma$ \\ \hline
 0.0000 & $1.2847 \times 10^{-9}$ & 1.7209 \\
$1.2372 \times 10^6$ & $8.2700 \times 10^{-8}$ & 1.4241 \\
$1.6870 \times 10^8$ & $1.2124 \times 10^{-6}$ & 1.2823 \\
$2.7977 \times 10^{11}$ & 4.2083 & 0.7109 \\
$3.4044 \times 10^{12}$ & $5.3155 \times 10^{-9}$ & 1.4210 \\
$7.6502 \times 10^{13}$ & $8.0820 \times 10^{-16}$ & 1.9121 \\
$2.0379 \times 10^{14}$ & $2.0848 \times 10^{-28}$ & 2.7918 \\
$6.0794 \times 10^{14}$ & $5.1814 \times 10^{-38}$ & 3.4415 \\
$1.0994 \times 10^{15}$ & $5.0414 \times 10^{-26}$ & 2.6445 \\
$1.8898 \times 10^{15}$ & $1.4171 \times 10^{-20}$ & 2.2878 \\
\hline 
\hline
\end{tabular}
\end{table*}

\end{document}